\documentclass[aps,prd,nofootinbib,twocolumn,showpacs,preprintnumbers,eqsecnum,notitlepage,10pt,floatfix]{revtex4-1}
\usepackage{color}
\usepackage{bm}
\usepackage{graphicx}
\usepackage{amsmath}
\usepackage{amssymb}
\usepackage{enumitem}
\usepackage{hyperref}
\usepackage{array}
\usepackage[english]{babel}
\usepackage{tensor}
\usepackage{comment}
\usepackage[normalem]{ulem}
\usepackage[dvipsnames]{xcolor}
\usepackage{amsfonts}
\usepackage{dcolumn}
\usepackage[caption=false]{subfig}

\usepackage{amsfonts}
\usepackage{dcolumn}

\newcommand{\newc}{\newcommand}
\newc{\tif}{\tilde{f}}
\newc{\tih}{\tilde{h}}
\newc{\tip}{\tilde{\phi}}
\newc{\tiA}{\tilde{A}}
\newc{\rM}{{\rm M}}

\newcommand{\mr}[1]{\mathrm{#1}}
\newcommand{\mL}[1]{\mathcal{#1}}

\newcommand{\ben}{\begin{eqnarray}}
\newcommand{\een}{\end{eqnarray}}
\newc{\be}{\begin{equation}}
\newc{\ee}{\end{equation}}
\newc{\ba}{\begin{eqnarray}}
\newc{\ea}{\end{eqnarray}}
\newc{\bea}{\begin{eqnarray*}}
\newc{\eea}{\end{eqnarray*}}
\newc{\D}{\partial}
\newc{\ie}{{\it i.e.} }
\newc{\eg}{{\it e.g.} }
\newc{\etc}{{\it etc.} }
\newc{\etal}{{\it et al.}}

\newc{\ra}{\rightarrow}
\newc{\lra}{\leftrightarrow}
\newc{\lsim}{\buildrel{<}\over{\sim}}
\newc{\gsim}{\buildrel{>}\over{\sim}}
\newc{\aP}{\alpha_{\rm P}}
\newc{\dphi}{\delta\phi}
\newc{\da}{\delta A}
\newc{\tp}{\dot{\phi}}
\newc{\Ve}{V_{\rm eff}}
\newc{\Vep}{V_{{\rm eff},\phi}}

\begin{document}

\title{Lunar Laser Ranging constraints on 
nonminimally coupled dark energy \\
and standard sirens}

\author{Shinji Tsujikawa}

\affiliation{
Department of Physics, Faculty of Science, 
Tokyo University of Science, 1-3, Kagurazaka,
Shinjuku-ku, Tokyo 162-8601, Japan
}

\begin{abstract}

In dark energy models where a scalar field $\phi$ 
is coupled to the Ricci scalar $R$ of the form 
$e^{-2Q (\phi-\phi_0)/M_{\rm pl}}R$, 
where $Q$ is a coupling constant, $\phi_0$ is today's value of $\phi$,  
and $M_{\rm pl}$ is the reduced Planck mass, 
we study how the recent Lunar Laser Ranging (LLR) experiment 
places constraints on the nonminimal coupling from 
the time variation of gravitational coupling. Besides 
a potential of the light scalar responsible for cosmic acceleration, 
we take a cubic Galileon term into account to suppress 
fifth forces in over-density regions of the Universe. 
Even if the scalar-matter interaction is screened by 
the Vainshtein mechanism, the time variation of 
gravitational coupling induced by the cosmological 
background field $\phi$ survives in the solar system.
For a small Galileon coupling constant $\beta_3$, there 
exists a kinetically driven 
$\phi$-matter-dominated-epoch ($\phi$MDE)
prior to cosmic acceleration. In this case, we 
obtain the stringent upper limit $Q \le 3.4 \times 10^{-3}$ from 
the LLR constraint. For a large $\beta_3$ without the 
$\phi$MDE, the coupling $Q$ is 
not particularly bounded from above, but the cosmological 
Vainshtein screening strongly suppresses the time 
variation of $\phi$ such that the dark energy 
equation of state $w_{\rm DE}$ reaches the value close 
to $-1$ at high redshifts. We study the modified gravitational wave
propagation induced by the nonminimal coupling 
to gravity and show that, under the LLR bound, the difference between 
the gravitational wave and luminosity distances 
does not exceed the order $10^{-5}$ over 
the redshift range $0<z<100$.
In dark energy models where the Vainshtein mechanism 
is at work 
through scalar derivative self-interactions, it is difficult to probe
the signature of nonminimal couplings from 
the observations of standard sirens.

\end{abstract}


\maketitle

\section{Introduction}

Since the first discovery of late-time cosmic acceleration 
by supernovae type Ia (SN Ia) in 1998 \cite{SN1,SN2}, 
the origin of this phenomenon has not been identified yet.
A scalar field $\phi$ is one of the simplest candidates 
for dark energy, whose potential energy \cite{quin} or 
nonlinear kinetic energy \cite{kes} can drive the acceleration. 
If we allow for the coupling between $\phi$ and the 
gravity sector, Horndeski theories \cite{Horndeski} are 
known as the most general scalar-tensor theories 
with second-order equations of motion \cite{Horn1,Horn2,Horn3}. 

Dark energy models based on Horndeski theories can be constrained 
not only by the observational data of SNIa, Cosmic Microwave 
Background (CMB) temperature anisotropies, Baryon Acoustic 
Oscillations (BAO) but also by the measurements of gravitational waves (GWs). 
The bound on the speed of GWs from gravitational Cherenkov 
radiation \cite{Moore} was used in Ref.~\cite{Kimura:2011qn} 
to place constraints on the Lagrangian of Horndeski theories. 
After the first discovery of the GW event GW150914 \cite{GW15}, 
the possibility for constraining modified gravity models from the 
measurements of GWs along with gamma-ray bursts was pointed 
out in Ref.~\cite{Lombriser:2015sxa}. 
{}From the Hulse-Taylor pulsar data, the speed of GWs $c_{t}$ 
was also constrained to be close to that of light $c$ 
at the level of $10^{-2}$ \cite{Jime}.

The GW170817 event from a neutron star merger \cite{GW170817} 
together with electromagnetic counterparts \cite{Goldstein} 
showed that the relative difference between $c_t$ and $c$
is less than the order 
$10^{-15}$. If we strictly demand that $c_t=c$ on the isotropic cosmological 
background, the allowed Horndeski Lagrangian is of the form 
$L=G_2(\phi, X)+G_3(\phi,X) \square \phi+G_4(\phi)R$, 
where $G_2, G_3$ are functions of $\phi$ and 
$X=-\partial_{\mu} \phi \partial^{\mu} \phi/2$, while
$G_4$ is a function of $\phi$ 
alone \cite{GW1,GW2,GW3,GW4,GW5}.
This includes the theories like 
quintessence \cite{quin}, 
k-essence \cite{kes}, 
cubic Galileons \cite{Nicolis,Galileon1,Galileon2,braiding}, 
Brans-Dicke (BD) theory \cite{Brans}, 
$f(R)$ gravity \cite{Bergmann,Ruz,Staro}, 
and nonminimally coupled theories with general functions 
$G_4(\phi)$ \cite{Damour1,Damour2,Amen99,Uzan,Chiba99,Bartolo99,Perrotta,Boi00,Gille,BD1}.

The original massless BD theory \cite{Brans} is equivalent 
to the Lagrangian $L=\left( 1-6Q^2 \right) F(\phi) X+
(M_{\rm pl}^2/2) F(\phi)R$ with 
$F(\phi)=e^{-2Q (\phi-\phi_0)/M_{\rm pl}}$, where 
the constant $Q$ is related to the so-called BD 
parameter $\omega_{\rm BD}$, as 
$2Q^2=1/(3+2\omega_{\rm BD})$ \cite{BD1}.
General Relativity (GR) is recovered in the limit 
$\omega_{\rm BD} \to \infty$, i.e., $Q \to 0$.
If we transform the action of BD theory to that in the Einstein 
frame, the constant $Q$ has a meaning of coupling between 
the scalar field and nonrelativistic 
matter \cite{DT10}.

The parametrized post-Newtonian (PPN) 
formalism \cite{Nord,Will71}
on the weak gravitational background shows that, in  
massless BD theory, one of the PPN parameters 
is given by 
$\gamma=(1+\omega_{\rm BD})/(2+\omega_{\rm BD})$ \cite{Will05}. 
The Cassini experiment measuring the time delay of light 
in the solar system placed the constraint 
$|\gamma-1| \le 2.3 \times 10^{-5}$ \cite{Will14}. 
This translates to the bound $\omega_{\rm BD} \ge 4.3 \times 10^4$, 
or equivalently, $|Q| \le 2.4 \times 10^{-3}$. 
For the coupling $|Q|>2.4 \times 10^{-3}$, one needs to 
resort to some mechanism for screening fifth forces mediated 
by the BD scalar field.

If the BD scalar has a massive potential in over-density 
regions of the Universe, the propagation of fifth forces 
can be suppressed under the chameleon mechanism \cite{chame1,chame2}.  
For example, metric $f(R)$ gravity corresponds to BD theory 
with $Q=-1/\sqrt{6}$ in the presence of a scalar potential 
of gravitational origin \cite{DT10,APT07}.
It is possible to design the form of $f(R)$ such that the scalar 
degree of freedom (scalaron) has a heavy mass in 
over-density regions, while realizing cosmic acceleration 
by a light scalar on Hubble scales \cite{fR1,fR2,fR3,fR4}.
However, this amounts to a fine-tuning of initial conditions of 
scalaron perturbations in the early Universe \cite{fR2,fR4,fR5}.
Moreover, unless the scalaron is nearly frozen until 
recently, the large coupling $|Q| \simeq 0.4$ leads to the 
significant enhancement of matter perturbations 
in the late Universe \cite{fR1,fR2,fR4,Tsuji07,Tsuji09}. 
For the compatibility of $f(R)$ models of late-time 
cosmic accelerationwith with observations,  
the deviation from GR is required to be 
very small and hence they are hardly distinguishable from the 
$\Lambda$-Cold-Dark-Matter ($\Lambda$CDM) model \cite{Lomb,Battye}.

There is yet another mechanism for screening fifth forces 
in local regions of the Universe based on nonlinear derivative 
self-interactions \cite{Vain}. 
A representative example is the cubic Galileon Lagrangian 
$X \square \phi$  \cite{Nicolis,Galileon1,Galileon2,braiding}, with which 
the Newtonian behavior is recovered inside the so-called Vainshtein radius 
$r_V$  \cite{Cede,Luty,Babichev,Burrage,Brax,Babi11,DKT12,KKY12,Kase13}
even with the coupling $|Q|>2.4 \times 10^{-3}$.
For uncoupled Galileons ($Q=0$) without the scalar potential, 
it is known that there exists a cosmological tracker solution finally 
approaching a de Sitter attractor \cite{DTGa,DTGa2} 
(see also Refs.~\cite{GS,Ali}).
Unfortunately, this dark energy model is in tension with 
the observational data of supernovae type Ia, CMB, 
BAO, and redshift-space 
distortions \cite{NDT10,AppleLin,Neveu,Barreira1,Barreira2,Renk,Peirone2}.  
For the nonminimally coupled light mass or massless Galileon with 
a potential, e.g., the linear potential $V(\phi)=m^3 \phi$, it is possible 
to realize the viable cosmic expansion history, while 
recovering the Newtonian behavior in the solar system \cite{Ali2,KTD}.

While the Vainshtein mechanism suppresses the scalar-matter interaction for the 
distance $r \ll r_V$, the gravitational coupling $G_{\rm N}$ in over-density regions contains 
time dependence of the dark energy field $\phi$ through 
the nonminimal coupling $F(\phi)$ \cite{Babi11,KKY12}. 
Then, $G_{\rm N}$ varies in time even inside the solar system. 
The LLR experiments of the earth-moon system measure 
the time variation $\dot{G}_{\rm N}/G_{\rm N}$, so it can 
be used to constrain nonminimally coupled dark energy models. 

{}From the LLR bound of $\dot{G}_{\rm N}/G_{\rm N}$ 
in 2004 \cite{Williams}, the time variation 
$\alpha_{\rM} \equiv \dot{F}/(HF)$ 
(where $H$ is the Hubble expansion rate) is in the range
$|\alpha_{\rM} (t_0)| \le 0.02$ today.
In 2011, Babichev {\it et al.} \cite{Babi11} used this bound
for nonminimally coupled cubic Galileons without the potential
and claimed that the time variation of the field is tightly constrained at low redshifts. 
We note that, besides this fact, the cubic Galileon without the potential 
is in tension with the observational data.  
On the other hand, the presence of potentials for nonminimally coupled 
Galileons allows the possibility for realizing viable cosmic 
expansion and growth histories, even with the LLR bound in 2004, 
see Figs.~4 and 5 in Ref.~\cite{KT18}.

The recent LLR experiments \cite{Hofmann} constrain the time variation 
$\dot{G}_{\rm N}/G_{\rm N}$ with the upper limit more stringent 
than before \cite{Williams}.
In particular, for $\alpha_{\rm M}>0$, the upper bound of  
$\dot{G}_{\rm N}/G_{\rm N}$ translates to 
$\alpha_{\rm M}(t_0) \le 7 \times 10^{-5}$ today, which is 
tighter than the bound $\alpha_{\rM} (t_0) \le 0.02$ 
by more than two orders of magnitude. 
This LLR bound in 2018 was used to constrain dark energy 
models based on nonlocal gravity \cite{Belga}. 
It remains to be seen how nonminimally coupled Galileons 
with the potential can be constrained with this new bound 
of $\alpha_{\rm M}(t_0)$.

In this paper, we exploit the new LLR bound to constrain 
nonminimally coupled dark energy models with 
the cubic self-interaction $\beta_3 M^{-3} X \square \phi$ 
and the potential $V(\phi)$ of light mass Galileons, where 
$\beta_3$ is dimensionless coupling constant and $M$ is 
a mass scale defined later in Eq.~(\ref{Massdef}).
We stress that our model is different from the nonminimally coupled 
cubic Galileon without the potential studied in Ref.~\cite{Babi11}, 
in that the scalar potential is the dominant source for late-time cosmic 
acceleration. The Galileon term can still play an important role 
for the scalar field dynamics in the early Universe. 
Moreover, we require that the propagation 
of fifth forces is suppressed in over-density regions. 
We perform a detailed analysis for the cosmological dynamics from 
the radiation era to today and put bounds on the coupling $Q$ 
by using the new LLR data.

For $|\beta_3| \ll 1$, there exists a so-called 
$\phi$-matter-dominated epoch ($\phi$MDE) \cite{Amenco} 
in the Jordan frame followed by the stage of cosmic acceleration. 
For the exponential potential $V(\phi)=V_0 e^{\lambda \phi/M_{\rm pl}}$, 
we place constraints on the allowed parameter space in 
the $(\lambda,Q)$ plane and derive the stringent limit  
$Q \le 3.4 \times 10^{-3}$ from the LLR constraint. 
This is almost close to the Cassini bound $Q \le 2.4 \times 10^{-3}$ 
obtained for massless BD theories without 
the Vainshtein screening. 
For $|\beta_3| \gg 1$, the coupling $Q$ is not particularly bounded from 
above due to the suppression of field kinetic energy under the 
cosmological Vainshtein screening. 
In this case, we show a new possibility for realizing 
the dark energy equation of state $w_{\rm DE}$ 
close to $-1$ from high redshifts to today even for the steep 
potential satisfying $\lambda>\sqrt{2}$.

In our dark energy theory the speed of GWs is equivalent to that 
of light, but the existence of nonminimal coupling $F(\phi)$ 
leads to the modified GW propagation through the existence 
of a nonvanishing term $\alpha_{\rM}$.
The possibility of using the difference between GW and luminosity 
distances to test for the running Planck mass 
was first pointed out in Ref.~\cite{Saltas:2014dha}.
The first forecasts of such constraints were 
given in Ref.~\cite{Lombriser:2015sxa}, 
which were followed by a sequence of papers after the 
direct detection of GWs \cite{Nishi17,Arai,Amendola17,Zhao,Belga17,Belga17d,Ezqu,Lagos}. 

In Ref.~\cite{Lombriser:2015sxa}, it was anticipated that the LLR bound 
on the running Planck mass may be beyond the reach of the constraint 
arising from standard sirens. This generally depends on the models of dark energy. 
For example, in nonlocal gravity models studied recently in Ref.~\cite{Belga},  
the difference between the GW distance  $d_{\rm GW}$ and luminosity distance $d_L$
is typically more than a few percent, which may be probed in future high-precision measurements. 
This reflects the fact that, in nonlocal gravity, the gravitational coupling deep inside the Hubble 
radius (wavelength $a/k \ll H^{-1}$) is very close to the Newton gravitational constant $G$, 
as $G_{\rm N}/G=1+{\cal O}((aH/k)^2)$ \cite{Nesseris,Belga}. 
Hence the nonlocal gravity models can pass
the new LLR bound in 2018, while leaving the sizable difference between 
$d_{\rm GW}$ and $d_L$. 

The nonminimal coupling $G_4(\phi)R$ gives rise to the effective 
gravitational coupling 
$G_{\rm N}$ different from that in nonlocal gravity.
Hence it deserves for studying whether the new LLR data leads to the constraint on the nonminimal coupling beyond or within the reach of future 
standard siren measurements.  
In this paper, we will compute the relative ratio between 
$d_{\rm GW}$ and $d_L$ for the aforementioned nonminimally 
coupled dark energy model. Under the LLR bound on the variation of $F(\phi)$, 
we show that the relative difference $d_{\rm GW}/d_L-1$ does not exceed 
the order $10^{-5}$ in the redshift range 
$0<z<100$. Thus, unlike nonlocal gravity, 
the LLR data allow only tiny deviations of $d_{\rm GW}$ from $d_L$ 
in nonminimally coupled theories, so 
it will be difficult to detect such difference without 
very high-precision distance measurements in future.

This paper is organized as follows.
In Sec.~\ref{sec:syspf}, we present our nonminimally coupled 
dark energy model and revisit how the cubic Galileon interaction screens the scalar-matter coupling under the
Vainshtein mechanism.
We then interpret the recent LLR bound in terms of today's 
value of $\alpha_{\rM}$. 
In Sec.~\ref{darksec}, we derive the background equations 
of motion on the flat Friedmann-Lema\^{i}tre-Robertson-Walker (FLRW) background and express them in 
autonomous forms. 
In Sec.~\ref{numesec}, we study the cosmological dynamics 
in the presence of exponential potential $V(\phi)=V_0 
e^{\lambda \phi/M_{\rm pl}}$ for unscreened 
($|\beta_3| \ll 1$) and screened ($|\beta_3| \gg 1$) cases 
after the radiation domination. 
We put constraints on the allowed parameter space from 
the recent LLR bound and discuss the evolution of 
$w_{\rm DE}$ and field density parameters. 
In Sec.~\ref{GWsec}, we investigate how much difference 
arises between $d_{\rm GW}(z)$ and $d_L(z)$ for the two 
different background cosmologies discussed in Sec.~\ref{numesec}.
Sec.~\ref{consec} is devoted to conclusions.

Unless otherwise stated, we use the natural unit where 
the speed of light $c$, the reduced Planck constant $\hbar$, 
and the Boltzmann constant $k_B$ 
are equivalent to 1.
 
\section{Nonminimally coupled theories and 
LLR constraints}
\label{sec:syspf}

We begin with a subclass of Horndeski theories 
given by the action  
\ba
\mL{S} 
&=& \int \mr{d}^4 x \sqrt{-g}  
\biggl[  \frac{M_{\rm{pl}}^2}{2} F(\phi) R
+\left( 1-6Q^2 \right) F(\phi) X \nonumber \\
& &\qquad \qquad \quad -V(\phi)
+\beta_3 M^{-3} X \square \phi
\biggr]+\mL{S}_m \,,
\label{action}
\ea
where $g$ is the determinant of metric tensor $g_{\mu \nu}$, 
$X=-\partial_{\mu} \phi \partial^{\mu} \phi/2$ is the kinetic  
energy of a scalar field $\phi$, and $F(\phi)$, $V(\phi)$ 
are functions of $\phi$. 
The couplings $Q$ and $\beta_3$ are dimensionless constants, 
while $M$ is a constant having a dimension of mass 
related to today's Hubble constant $H_0$ as
\be
M=\left( M_{\rm pl} H_0^2 \right)^{1/3}\,,
\label{Massdef}
\ee
which is of order $10^{-22}$~GeV. 
The mass scale (\ref{Massdef}), which translates to the frequency 
$f \sim 100$~Hz, corresponds to the typical strong coupling scale of 
theories containing derivative self-interactions 
like $X \square \phi$ \cite{Cede,Luty}.

We assume that the effective field theory of dark energy is valid up 
to the mass scale $M \sim 10^{-22}$~GeV. 
In other words, we resort to the action (\ref{action}) 
for studying the physics on scales larger than $\sim 10^6$~m. 
This includes the dynamics of late-time cosmic acceleration 
($\sim 10^{26}$~m) and the earth-moon local 
system of LLR experiments ($\sim 10^8$~m).
Below the length scale $10^6$~m, some ultraviolet effects may 
come into play to approach the General Relativistic behavior.
Indeed, the possibility of recovering the GW value $c_t=1$ above 
the frequency $f \sim 100$~Hz was discussed in Ref.~\cite{deRham:2018red}.
This critical frequency is of the same order as the GW frequency 
observed by LIGO/Virgo \cite{GW170817}, 
so there may be more general Horndeski theories realizing 
$c_t$ very close to 1 for the frequency $f\geq 100$~Hz even if 
the deviation of $c_t$ from 1 is large on cosmological scales.
In this paper we do not pursue such a possibility, but we focus 
on the theory given by the action (\ref{action}) in which $c_t=1$ 
for any scales of interest.

The nonminimal coupling $F(\phi)$ is chosen to be 
of the form:
\be
F(\phi)=e^{-2Q (\phi-\phi_0)/M_{\rm pl}}\,,
\label{Fphi}
\ee
where $\phi_0$ is today's value of $\phi$ and hence
$F(\phi_0)=1$. 
We assume that the matter sector, which is described 
by the action $\mL{S}_m$ with the density $\rho_m$, 
is minimally coupled to gravity. 
The scalar field mediates fifth forces with the matter sector 
through the direct gravitational interaction characterized by 
the coupling $Q$.

If $\beta_3=0$, then the theories given by the action (\ref{action}) 
are equivalent to BD theories \cite{Brans} 
with the scalar potential $V(\phi)$.
Indeed, by setting $\chi=F(\phi)$, the Lagrangian 
in the action (\ref{action}) reduces to 
$L=\chi R/2-\omega_{\rm BD} 
\partial_{\mu} \chi \partial^{\mu} \chi/(2\chi)-V(\phi (\chi))$ 
in the unit $M_{\rm pl}=1$, where $\omega_{\rm BD}$ 
is the BD parameter related to $Q$ according to 
$3+2\omega_{\rm BD}=1/(2Q^2)$ \cite{BD1}. 
In the original massless BD theories with $V(\phi)=0$, 
the coupling strength is constrained to be 
$|Q| \le 2.4 \times 10^{-3}$ from the Cassini experiment \cite{Will14}.

For the coupling $|Q|>2.4 \times 10^{-3}$, we require the 
existence of scalar potential $V(\phi)$ or field derivative interaction 
$X \square \phi$ to screen fifth forces in the solar system.
In the former case, the chameleon mechanism \cite{chame1,chame2} 
can be at work for the potential having a large mass in regions of the high density. 
One of such examples is $f(R)$ gravity, in which the scalar 
potential of gravitational origin arises with the coupling 
$Q=-1/\sqrt{6}$ \cite{DT10}. 
In $f(R)$ models of late-time cosmic acceleration accommodating 
the chameleon mechanism in over-density regions, 
the functional form of $f(R)$ needs to be designed such that
the scalaron mass $M_{\phi}$ grows very rapidly
toward the asymptotic past \cite{fR1,fR2,fR3,fR4}.
This causes the fine-tuning problem of initial conditions of perturbations 
associated with the oscillating mode induced 
by the heavy mass \cite{fR2,fR4,fR5}.

Instead of resorting to the chameleon mechanism with 
a very massive scalar in over-density regions, 
we consider the Galileon self-interaction $X \square \phi$ 
to suppress fifth forces under the Vainshtein 
mechanism \cite{Vain}. 
The scalar potential $V(\phi)$ of a light scalar is also 
taken into account as a source for the cosmic acceleration. 
Defining the dimensionless quantity
\be
\lambda \equiv \frac{M_{\rm pl}}{V} \frac{{\rm d}V}{{\rm d}\phi}\,,
\ee
the condition for cosmic acceleration in the absence of Galileon 
interactions and matter is given by 
$|\lambda|<\sqrt{2}$ \cite{CLW,CST}.
The existence of Galileons can modify this structure, 
but we focus on the case in which the condition
\be
\left| \lambda \right| \le {\cal O}(1)
\label{lamra}
\ee
is satisfied during the cosmic expansion history from the past 
to today. The coupling strength $|Q|$ exceeding the order 1 leads to the 
strong enhancement of matter density perturbations 
incompatible with observations in large-scale structures \cite{BD1}, 
so we consider the coupling 
\be
|Q| \le {\cal O}(0.1)\,,
\label{Qra}
\ee
in the following discussion.

The original Galileon theory \cite{Nicolis} has the linear 
potential $V(\phi)=m^3 \phi$ with $Q=0$, in which case 
the resulting field equation of motion
respects the Galilean symmetry in Minkowski space-time. 
This potential corresponds to a massless scalar 
with $\lambda=M_{\rm pl}/\phi$, 
so the condition (\ref{lamra}) translates to 
$\phi \ge M_{\rm pl}$. 
For $Q \neq 0$, the cosmological dynamics with 
the linear potential was studied in 
Ref.~\cite{KTD}. In this case, today's cosmic acceleration 
is followed by the collapsing Universe after the field enters 
the region $V(\phi)<0$. 

The constant $\lambda$ corresponds to the exponential potential 
$V(\phi)=V_0 e^{\lambda \phi/M_{\rm pl}}$. 
In this case, the scalar mass squared $M_{\phi}^2 \equiv {\rm d}^2 V/{\rm d}\phi^2$ 
is given by $M_{\phi}^2=\lambda^2 V/M_{\rm pl}^2$. 
Since the potential energy $V$ is the dominant contribution to 
today's energy density of the Universe, we have 
$V \lesssim M_{\rm pl}^2 H^2$, where $H$ is the Hubble expansion rate in the past (redshift $z \geq 0$).
Then, under the condition (\ref{lamra}), it follows that $M_{\phi}^2 \lesssim \lambda^2 H^2 \lesssim H^2$.
This property also holds for the potential with 
a time-varying $\lambda$ in the range (\ref{lamra}). 
For the light scalar whose today's mass $M_{\phi}$ is smaller 
than $H_0$, the effect of $M_{\phi}$ on the scalar-field 
equation can be ignored to study the Vainshtein mechanism 
in regions of the high density.
In other words, the chameleon mechanism does not 
come into play for screening fifth forces.

\subsection{Vainshtein screening}

The behavior of scalar and gravitational fields around 
a spherically symmetric over-density on a cosmological background was already studied in 
Refs.~\cite{KKY12,Babi11}, so we briefly review it in the following.
Let us consider the following perturbed metric in the Newtonian gauge: 
\be
{\rm d}s^2=- \left( 1+2\Psi \right) {\rm d}t^2
+\left( 1+2\Phi \right) a^2(t) \delta_{ij}{\rm d}x^i {\rm d}x^j\,,
\ee
where $a(t)$ is the time-dependent scale factor, $\Psi$ and $\Phi$ 
are gravitational potentials depending $t$ and 
the radial coordinate $r=a(t)\sqrt{ \delta_{ij} x^i x^j}$. 
The scalar field and matter density on the homogenous cosmological 
background are given by $\bar{\phi}(t)$ and $\bar{\rho}_m(t)$, respectively.
The existence of a compact object gives rise to the perturbations 
$\chi (t, r)$ and $\delta \rho_m (t, r)$ in $\phi$ and $\rho_m$, 
such that $\phi=\bar{\phi}(t)+\chi (t, r)$ and 
$\rho_m=\bar{\rho}_m(t)+\delta \rho_m (t, r)$. 

We are interested in solutions deep inside today's 
Hubble radius, $r \ll H_0^{-1}$. 
Hence we neglect time derivatives of perturbed quantities, 
while keeping spatial derivatives. 
The radial dependence of the derivative 
$\partial \chi/\partial r$ changes around the Vainshtein radius 
$r_V$, which is estimated as \cite{KKY12,DKT12}
\be
r_V \simeq \left( \frac{|\beta_3 Q| M_{\rm pl} r_g}{M^3}
\right)^{1/3}= \left( |\beta_3 Q| r_g H_0^{-2} \right)^{1/3}\,,
\label{rV}
\ee
where 
\be
r_g=M_{\rm pl}^{-2} \int_0^r \delta \rho_m\,
\tilde{r}^2 {\rm d}\tilde{r}
\label{rg}
\ee
is the Schwarzschild radius of the source. 
For $r \gg r_V$ the field derivative has the dependence 
$\partial \chi/\partial r \propto r^{-2}$, while, for $r \ll r_V$, 
$\partial \chi/\partial r \propto r^{-1/2}$.
In the latter regime, the nonlinear effect arising from the cubic 
Galileon self-interaction suppresses the propagation of fifth forces 
induced by the coupling $Q$.
Indeed, for $r \ll r_V$, the gravitational potentials 
are given by \cite{KKY12,DKT12}
\ba
\Psi &\simeq& -\frac{r_g}{2rF}
\left[ 1+{\cal O}(1)\,Q^2 \left( \frac{r}{r_V} 
\right)^{3/2} \right]\,,\label{Psilo}\\
\Phi &\simeq& \frac{r_g}{2rF}
\left[ 1+{\cal O}(1)\,Q^2 \left( \frac{r}{r_V} 
\right)^{3/2} \right]\,.\label{Philo}
\ea
Since the value of $F$ today (cosmic time $t_0$) is 
equivalent to 1 in our theory, the Newtonian 
behavior ($-\Psi=\Phi=r_g/(2r)$) is 
recovered for $r \ll r_V$.
As long as $r_V$ is much larger than 
the solar-system scale ($\sim 10^{15}$\,cm), the model is 
consistent with solar-system tests of gravity. 
Since $(r_gH_0^{-2})^{1/3} \simeq 3 \times 10^{20}$~cm for 
the Sun, this condition translates to 
\be
\left| \beta_3 Q \right| \gg 10^{-17}\,.
\label{local}
\ee
When $|Q|$ is of order $10^{-2}$, for example, 
the coupling $\beta_3$ needs to 
be in the range $|\beta_3| \gg 10^{-15}$.

\subsection{LLR constraints}

{}From Eq.~(\ref{Psilo})-(\ref{Philo}) with Eq.~(\ref{rg}), the leading-order gravitational potentials deep inside the Vainshtein radius can be expressed as 
\be
-\Psi \simeq \Phi \simeq  \frac{G_{\rm N} \delta {\cal M}}{r}\,,
\ee
where $ \delta {\cal M}=4\pi \int_0^r \delta \rho_m \tilde{r}^2 {\rm d} \tilde{r}$, and 
$G_{\rm N}$ is the measured gravitational coupling given by 
\be
G_{\rm N}=\frac{1}{8 \pi M_{\rm pl}^2 F(\phi(t))}\,,
\label{GN}
\ee
where we omitted the bar from the background value of $\phi$.
Here the background field $\phi(t)$ is a cosmological scalar 
driving the late-time cosmic acceleration.
Since we are considering over-density regions on the cosmological 
background, the homogenous value $\phi(t)$ survives even in 
the local Universe.
The dark energy scalar field $\phi(t)$ changes in time, so 
this leads to the time variation of $G_{\rm N}$. 
This fact was first recognized in Ref.~\cite{Babi11} and it was 
proved in Ref.~\cite{KKY12} in full Horndeski theories.

The effective gravitational coupling (\ref{GN}) 
is valid for a light scalar field operated by the Vainshtein 
mechanism in over-density regions. 
Here, the light scalar means that the slope of field potential 
$V(\phi)$ satisfies the condition (\ref{lamra}). 
For the potential of a massive scalar violating this condition in regions 
of the high density (as in $f(R)$ dark energy models), 
the chameleon mechanism can be at work 
to suppress the gravitational coupling with matter
in a way different from Eqs.~(\ref{Psilo})-(\ref{Philo}).
As we already mentioned, we do not consider such a massive 
scalar field in this paper.

For the cubic derivative self-interaction we chose the Galileon 
coupling $X \square \phi$, but this can be generalized to the 
derivative coupling $X^n \square \phi$ with $n>1$. 
In such cases, the second terms on the right hand sides 
of (\ref{Psilo}) and (\ref{Philo}) are modified to 
${\cal O}(1) Q^2 (r/r_V)^{2-1/(2n)}$, which is much smaller 
than 1 deep inside the Vainshtein radius. 
Then the local gravitational coupling reduces to 
the form (\ref{GN}), so the property of $G_{\rm N}$ 
induced by the time-dependent background scalar field 
$\phi(t)$ is similar to that of cubic Galileons. 
For the models in which derivative field self-interactions 
are not employed to screen fifth forces in over-density regions, 
e.g., chameleons and nonlocal gravity,
the expression of $G_{\rm N}$ is generally different from 
that discussed above.

{}From the recent LLR experiment, the variation of $G_{\rm N}$ 
is constrained to be  \cite{Hofmann}
\be
\frac{\dot{G}_{\rm N}}{G_{\rm N}}=\left( 7.1 \pm 7.6 \right) 
\times 10^{-14}~{\rm yr}^{-1}\,,
\label{Gbou}
\ee
where a dot represents the derivative with respect to $t$.
This improves the previous bound 
$\dot{G}_{\rm N}/G_{\rm N}=(4 \pm 9) \times 10^{-13}$~yr$^{-1}$ 
\cite{Williams}. 
Using the value $H_0=100~h$~km s$^{-1}$ Mpc$^{-1}
=(9.77775~{\rm Gyr})^{-1}h$, 
the bound (\ref{Gbou}) translates to \cite{Belga}
\be
\frac{\dot{G}_{\rm N}}{H_0G_{\rm N}}=
\left( 0.99 \pm 1.06 \right) \times 10^{-3}  
\left( \frac{0.7}{h} \right)\,.
\label{Gbou2}
\ee
We define the following quantity,
\be
\alpha_{\rM} \equiv \frac{\dot{F}}{HF}
=-\frac{2Q \dot{\phi}}{M_{\rm pl}H}\,,
\label{alM}
\ee
which was used in the context of effective field theory of 
dark energy \cite{Bellini}. 
Since $\alpha_{\rM}$ is related to the variation 
of $G_{\rm N}$, as 
$\alpha_{\rM}=-\dot{G}_{\rm N}/(H G_{\rm N})$, 
the bound (\ref{Gbou2}) can be expressed as
\be
-2.05 \times 10^{-3} \left( \frac{0.7}{h} \right) \le
\alpha_{\rM} (t_0) \le 0.07 \times 10^{-3} \left( \frac{0.7}{h} \right)\,.
\label{aMcon}
\ee
If $\alpha_{\rM}>0$, i.e., for decreasing $G_{\rm N}$ in time, 
the upper bound is especially stringent: 
$\alpha_{\rm M}(t_0) \le 7 \times 10^{-5}$ for $h=0.7$.
Even when $\alpha_{\rM}<0$, the upper limit of 
$|\alpha_{\rM} (t_0)|$ is of the order $10^{-3}$. 
They are smaller than the previous bound
$|\alpha_{\rM} (t_0)| \le 0.02$ \cite{Babi11}
by more than one order of magnitude.

\section{Dynamical system}
\label{darksec}

We study the background cosmology for theories given 
by the action (\ref{action}) and discuss how the coupling 
$Q$ is constrained from the LLR bound (\ref{aMcon}). 
We consider the flat FLRW background described by the line element 
${\rm d}s^2=-{\rm d}t^2+a^2(t) \delta_{ij}{\rm d} x^i {\rm d}x^j$.
For the matter action $\mL{S}_m$, we take nonrelativistic 
matter (density $\rho_m$ with vanishing pressure) and 
radiation (density $\rho_r$ and pressure $P_r=\rho_r/3$) 
into account. 
Then, the Hamiltonian and momentum constraints 
lead to \cite{Horn2,KT18}:
\ba
& &
3M_{\rm pl}^2 H^2=\rho_{\rm DE}+\rho_m+\rho_r\,,
\label{back1} \\
& &
2M_{\rm pl}^2 \dot{H}=-\rho_{\rm DE}-P_{\rm DE}
-\rho_m-\frac{4}{3} \rho_r\,,
\label{back2} 
\ea
where $H=\dot{a}/a$, and 
$\rho_{\rm DE}$ and $P_{\rm DE}$ are the density 
and pressure of dark energy, defined, 
respectively, by 
\ba
\rho_{\rm DE} 
&=& 3M_{\rm pl}^2 H^2 \left( 1-F \right)
+\frac{F}{2} (1-6Q^2) \dot{\phi}^2 \nonumber \\
& &+6FQ H M_{\rm pl} \dot{\phi}+V
-3 \beta_3 M^{-3} H \dot{\phi}^3\,,
\label{rhode} \\
P_{\rm DE} 
&=& -M_{\rm pl}^2\left( 2\dot{H}+3H^2 \right) 
\left( 1-F \right)+\frac{F}{2} (1+2Q^2) \dot{\phi}^2 
\nonumber \\
& &-2FQM_{\rm pl} \left( \ddot{\phi}+2H \dot{\phi} 
\right)-V+\beta_3 M^{-3} \dot{\phi}^2 \ddot{\phi}\,.
\label{Pde}
\ea
Besides the matter continuity equations $\dot{\rho}_m+3H \rho_m=0$ 
and $\dot{\rho}_r+4H \rho_r=0$, the dark sector obeys 
\be
\dot{\rho}_{\rm DE}+3H \left( \rho_{\rm DE}
+P_{\rm DE} \right)=0\,.
\label{back3}
\ee
The dark energy equation of state is defined by 
\be
w_{\rm DE} \equiv \frac{P_{\rm DE}}
{\rho_{\rm DE}}\,.
\label{wdedef}
\ee
In nonminimally coupled theories the first terms on the right 
hand sides of Eqs.~(\ref{rhode}) and (\ref{Pde}) are different from 
0 in the past due to the property $F \neq 1$.

To study the background cosmological dynamics, 
we introduce the following density parameters,
\ba
& &
\Omega_K \equiv \frac{\dot{\phi}^2}{6M_{\rm pl}^2H^2},
\qquad
\Omega_{V} \equiv 
\frac{V(\phi)}{3M_{\rm pl}^2 H^2 F}, \nonumber \\
& &
\Omega_{G_3} \equiv 
-\frac{\beta_3 \dot{\phi}^3}{M_{\rm pl}^2 M^3 H F},
\qquad
\Omega_r \equiv \frac{\rho_r}{3M_{\rm pl}^2 H^2 F}\,.
\label{moDdi}
\ea
We consider the case in which $\Omega_{G_3}$ is positive
in the expanding Universe ($H>0$),  
which amounts to the condition
\be
\beta_3 \dot{\phi}<0\,.
\ee
We also define the quantity 
\be
x \equiv \frac{\dot{\phi}}{\sqrt{6}M_{\rm pl}H}\,,
\ee
which is related to $\Omega_K$ and $\alpha_{\rM}$, as
\be
\Omega_K=x^2\,,\qquad
\alpha_{\rM}=-2\sqrt{6} Qx\,.
\ee
We can express Eq.~(\ref{back1}) in the form:
\be
\Omega_m \equiv \frac{\rho_m}{3M_{\rm pl}^2 H^2 F}=
1-\Omega_{\rm DE}-\Omega_r\,,
\label{Omem2}
\ee
where $\Omega_{\rm DE}$ is defined by 
\be
\Omega_{\rm DE} \equiv \left(1-6Q^2 \right) \Omega_K
-\alpha_{\rM}+\Omega_V+\Omega_{G_3}\,.
\ee

{}From Eqs.~(\ref{back2}) and (\ref{back3}), it follows that 
\begin{widetext}
\ba
h \equiv \frac{\dot{H}}{H^2}
&=&
-\frac{1}{{\cal D}}
\left[ \Omega_{G_3} ( 6+2\Omega_r
-6\Omega_{V}+3\Omega_{G_3}
-\alpha_{\rM}+\sqrt{6}\Omega_V \lambda x )
+2\Omega_K \{ 3+\Omega_r-3\Omega_{V}+6\Omega_{G_3}
+6\lambda Q \Omega_{V} \right. \nonumber \\
& &\qquad \left. +6Q^2 (1-\Omega_r+3\Omega_{V}
-2\Omega_{G_3}) \} -\alpha_{\rM} \Omega_K (1-6Q^2) 
(2-\Omega_{G_3})+6\Omega_K^2(1-8Q^2+12Q^4) \right]\,,\label{hD}\\
\epsilon_{\phi} \equiv \frac{\ddot{\phi}}{H \dot{\phi}}
&=& \frac{1}{{\cal D}}
[\Omega_{G_3} (\Omega_r-3-3\Omega_{V} )
-\alpha_{\rM} (\Omega_r-1-3\Omega_{V}
-2\Omega_{G_3})-2\sqrt{6} \Omega_V \lambda x \nonumber \\
& &\quad
-3 \Omega_K \{ 4(1-2Q^2)
-\Omega_{G_3}(1+2Q^2) \}
-\alpha_{\rM} \Omega_K (5-6Q^2)]\,,
\label{epD}
\ea
\end{widetext}
where
\be
{\cal D}=\Omega_{G_3} 
\left( 4-2\alpha_{\rM}+\Omega_{G_3} \right)
+4\Omega_K\,.
\ee
The condition for cosmic acceleration to occur is that 
the effective equation of state,
\be
w_{\rm eff} \equiv -1-\frac{2}{3}h\,,
\ee
is smaller than $-1/3$.

The dimensionless variables $x$, $\Omega_{V}$, $\Omega_{G_3}$, and $\Omega_r$ obey the differential equations, 
\ba
x' &=& x \left( \epsilon_{\phi}- h \right)\,,
\label{auto1} \\
\Omega_{V}' &=& -\Omega_{V} 
\left( \alpha_{\rM}-\sqrt{6}\lambda x+2h \right)\,,
\label{auto2} \\ 
\Omega_{G_3}' &=& -\Omega_{G_3} \left( 
\alpha_{\rM}-3\epsilon_{\phi}+h \right)\,, 
\label{auto3} \\ 
\Omega_r' &=& -\Omega_r 
\left( \alpha_{\rM}+4+2h \right)\,,
\label{auto4}
\ea
respectively, where a prime represents a derivative with respect to ${\cal N}=\ln a$.
The dark energy equation of state (\ref{wdedef}) 
is expressed as
\begin{widetext}
\be
w_{\rm DE}=-\frac{3+2h-[3+2h+3(1+2Q^2)\Omega_K
-3\Omega_{V}
+\alpha_{\rM} (2+\epsilon_{\phi})
-\epsilon_{\phi}\Omega_{G_3}]F}
{3-3[1+(6Q^2-1)\Omega_K
-\Omega_{V}+\alpha_{\rM}-\Omega_{G_3}]F}\,.
\label{wdeD}
\ee
\end{widetext}
The dimensionless field $y \equiv \phi/M_{\rm pl}$
obeys 
\be
y'=\sqrt{6}x\,.
\label{dy}
\ee
Once the potential $V(\phi)$ is specified, 
the cosmological dynamics is known by solving 
Eqs.~(\ref{auto1})-(\ref{auto4}) and (\ref{dy}) 
for given initial conditions of $x$, $\Omega_V$, 
$\Omega_{G_3}$, $\Omega_r$, and $y$.

For the theory (\ref{action}), the propagation 
speed squared of GWs is equivalent to 1 \cite{Horn2,DT12}. 
The tensor ghost is absent for $F(\phi)>0$, 
which is satisfied for the choice (\ref{Fphi}). 
For scalar perturbations, the conditions for avoiding 
ghosts and Laplacian instabilities are given, respectively, by 
\ba
\hspace{-0.8cm}
q_s &\equiv& \Omega_{G_3} \left( 4+ \Omega_{G_3}
-2\alpha_{\rM} \right)+4\Omega_K >0,\label{Qs}\\
\hspace{-0.8cm}
c_s^2 &\equiv& \frac{ \Omega_{G_3}[4(2+\epsilon_{\phi})
-\Omega_{G_3}-2\alpha_{\rM} ]+12\Omega_K}
{3\Omega_{G_3} \left( 4+ \Omega_{G_3}
-2\alpha_{\rM} \right)+12\Omega_K}>0\,.
\label{cs}
\ea
In Sec.~\ref{numesec}, we will discuss whether these conditions are satisfied during the cosmological evolution from the radiation-dominated epoch to today.

\section{Cosmological dynamics}
\label{numesec}

In this section, we study the cosmological dynamics 
for constant $\lambda$, i.e., the exponential potential, 
\be
V(\phi)=V_0 e^{\lambda \phi/M_{\rm pl}}\,.
\label{exp}
\ee
In this case, the dynamical system given by 
Eqs.~(\ref{auto1})-(\ref{auto4}) is closed. 
As long as $\lambda$ slowly varies in time in the range 
(\ref{lamra}), the cosmological evolution is similar to 
that discussed below. 

In over-density regions of the Universe, the operation of Vainshtein mechanism means that the cubic Galileon term 
$X \square \phi$ dominates over other field Lagrangians. 
In the cosmological context, this amounts to the dominance of 
$\Omega_{G_3}$ over $\Omega_K$ and $\Omega_V$ 
in the early epoch.
Let us consider the case in which the conditions 
\be
\{ \Omega_K, \Omega_V \} \ll \Omega_{G_3} \ll 1\,,
\qquad \left| \alpha_{\rM} \right| \ll 1
\label{radcon}
\ee
are satisfied during the radiation-dominated epoch
(in which $\Omega_r$ is close to 1). {}From Eqs.~(\ref{hD}) 
and (\ref{epD}), we then have $h \simeq -2$ and 
\be
\epsilon_{\phi} \simeq 
-\frac{1}{2}+\epsilon_{\alpha}\,,\qquad 
\epsilon_{\alpha} \equiv \frac{\alpha_{\rM}}{4 \Omega_{G_3}}
 \left( 1-\Omega_r \right)\,. 
\label{hepes}
\ee
Since $\Omega_r$ starts to deviate from 1 in the 
late radiation era, the term $\epsilon_{\alpha}$ is 
not necessarily negligible 
relative to $-1/2$ for $|\alpha_{\rM}| \gg \Omega_{G_3}$.
On using Eqs.~(\ref{auto1}), (\ref{auto3}), and (\ref{auto4}), the quantity $\epsilon_{\alpha}$ 
obeys the differential equation, 
\be
\epsilon_{\alpha}' \simeq 6Q^2 
\frac{\Omega_K}{\Omega_{G_3}}
+2 \epsilon_{\alpha}
 \left(  1- \epsilon_{\alpha} \right)\,.
\label{epp}
\ee
Under the condition $\Omega_{G_3} \gg \Omega_K$, 
the first term on the right hand side of Eq.~(\ref{epp}) is much smaller than 1. 
Ignoring this term and solving the 
differential equation $\epsilon_{\alpha}' \simeq 
2 \epsilon_{\alpha} \left(  1- \epsilon_{\alpha} \right)$
for $\epsilon_{\alpha}$, it follows that 
\be
\epsilon_{\alpha}=\left[1+\frac{a_i^2}{a^2}
\frac{1-\epsilon_{\alpha}^{(i)}}{\epsilon_{\alpha}^{(i)}} \right]^{-1}\,,
\label{epso}
\ee
where $\epsilon_{\alpha}^{(i)}$ is the initial value of 
$\epsilon_{\alpha}$ at $a=a_i$. 
In the limit $a \to \infty$, $\epsilon_{\alpha}$ 
asymptotically approaches 1.

If the condition $|\alpha_{\rM}| \gg \Omega_{G_3}$ is 
initially satisfied, $|\epsilon_{\alpha}^{(i)}|$ 
can be as large as the order 1. 
Then, $\epsilon_{\phi}$ soon approaches 
the asymptotic value
\be
\epsilon_{\phi} \to \frac{1}{2}\,,
\label{epphies}
\ee
during the radiation era. In this regime, 
the field density parameters and $|\alpha_{\rM}|$ 
grow as
\be
\Omega_K \propto a^5\,,\quad 
\Omega_V \propto a^4\,,\quad 
\Omega_{G_3} \propto a^{7/2}\,,\quad 
|\alpha_{\rM}| \propto a^{5/2}\,.
\label{Omeevo}
\ee
This shows that, even if $\Omega_{G_3} \gg \Omega_K$ initially, it is possible for $\Omega_K$ to 
catch up with $\Omega_{G_3}$. 
If this catch up occurs by the end of radiation era, 
we have $\Omega_{G_3}<\Omega_{K}$ 
at the onset of matter dominance.

If $|\alpha_{\rM}| \ll \Omega_{G_3}$ initially, 
i.e., $|\epsilon_{\alpha}^{(i)}| \ll 1$, 
there is the stage of radiation era in which 
the quantity $\epsilon_{\phi}$ is close to $-1/2$. 
On using Eqs.~(\ref{auto1})-(\ref{auto3}) in this epoch, 
the field density parameters and $|\alpha_{\rM}|$ 
evolve as
\be
\Omega_K \propto a^3\,,\quad 
\Omega_V \propto a^4\,,\quad 
\Omega_{G_3} \propto a^{1/2}\,,\quad 
|\alpha_{\rM}| \propto a^{3/2}\,,
\label{den1}
\ee
so that $|\alpha_{\rM}|$ grows faster than $\Omega_{G_3}$. 
If $|\alpha_{\rM}|$ exceeds $\Omega_{G_3}$ 
during the radiation era, the solutions enter the regime characterized by Eqs.~(\ref{epphies}) and (\ref{Omeevo}). 
Although $\Omega_K$ grows faster than 
$\Omega_{G_3}$ in the two regimes explained above, 
it can happen that the inequality 
$\Omega_{G_3}>\Omega_K$ still holds at the beginning 
of matter era for $\Omega_{G_3}$ initially much larger 
than $|\alpha_{\rM}|$ and $\Omega_K$.

The above discussion shows that there are two qualitatively 
different cases depending on the values of 
$\Omega_{G_3}$ and $\Omega_K$ 
at the onset of matter dominance.
The first is the case in which $\Omega_K$
dominates over $\Omega_{G_3}$, i.e., 
\be
{\rm (i)}~~\Omega_{G_3} \ll \Omega_{K}
\quad ({\rm unscreened})\,.
\ee
Under this condition, there exists 
the $\phi$MDE in which 
the field kinetic energy is not screened by the  
Galileon term.

The second is the case in which the condition 
\be
{\rm (ii)}~~\Omega_{G_3} \gg \Omega_{K}
\quad ({\rm screened})
\ee
is satisfied after the end of radiation era.
This corresponds to the situation in which  
the cosmological Vainshtein screening 
is sufficiently efficient to suppress the time variation 
of $\phi$ throughout the evolution 
from the radiation era to today.
In the following, we study these two different cases 
in turn.

We note that, under the conditions (\ref{radcon}),
the dark energy equation of state (\ref{wdeD})
during the radiation dominance can be estimated as
\be
w_{\rm DE} \simeq w_{\rm eff} \simeq 
\frac{1}{3}\,,
\label{wdera}
\ee
irrespective of the two asymptotic values of  
$\epsilon_{\phi}~(=\pm 1/2)$ explained above.

\subsection{Unscreened late-time cosmology with 
the $\phi$MDE}
\label{unsec}

Let us first study the cosmological dynamics for the case 
(i), i.e., $\Omega_{G_3} \ll \Omega_K$ after the onset 
of matte era. In this case, the coupling $\beta_3$ 
is in the range
\be
\left| \beta_3 \right| \ll 1\,.
\ee
To derive fixed points of the dynamical system, 
we take the limit 
$\Omega_{G_3} \to 0$ in the autonomous 
Eqs.~(\ref{auto1})-(\ref{auto4}). 
For $Q \neq 0$, the standard matter era is replaced 
by the $\phi$MDE characterized by the fixed point  
\be
({\rm a})~\left( x, \Omega_{V}, \Omega_{G_3}, \Omega_{r} 
\right) =\left( -\frac{\sqrt{6}Q}{3(1-2Q^2)},0,0,0 \right)\,,
\label{phiMDE}
\ee
with 
\ba
& &
\Omega_m=\frac{3-2Q^2}{3(1-2Q^2)^2}\,,\quad
w_{\rm eff}=\frac{4Q^2}{3(1-2Q^2)}\,, \nonumber \\
& &
w_{\rm DE}=\frac{4Q^2(1-2Q^2)}{3(1-F)-2(6-F)Q^2+12Q^4}\,.
\label{phiMDE2}
\ea
The $\phi$MDE was originally found for coupled quintessence in the 
Einstein frame \cite{Amenco}. 
This corresponds to the kinetically driven stage 
in which $\Omega_K=2Q^2/[3(1-2Q^2)^2]$ dominates 
over $\Omega_{G_3}$. 
On the fixed point (a), the parameter $\alpha_{\rM}$ is 
given by 
\be
\alpha_{\rM}^{(\rm a)}=\frac{4Q^2}{1-2Q^2}\,,
\label{aMa}
\ee
and hence $\alpha_{\rM}^{(\rm a)}>0$ for $Q^2<1/2$. 
The positivity of $\alpha_{\rM}^{(\rm a)}$ means that 
\be
Qx_{(\rm a)}<0\,,
\label{Qx}
\ee
where $x_{(\rm a)}$ is the value of $x$ on the $\phi$MDE.

After $\Omega_K$ exceeds $\Omega_{G_3}$ by the end 
of radiation era, the solutions are naturally followed 
by the $\phi$MDE 
in which the cosmological Vainshtein screening is no 
longer effective.
While $\Omega_K$ is constant during the $\phi$MDE, 
the other field density parameters evolve as
\be
\Omega_V \propto a^{\frac{3-2Q \lambda-6Q^2}{1-2Q^2}}\,,\quad 
\Omega_{G_3} \propto 
a^{-\frac{3+2Q^2}{1-2Q^2}}\,.
\label{OmeVG}
\ee
For $|Q \lambda| \ll 1$ and $Q^2 \ll 1$, $\Omega_V$ grows 
in proportion to $a^3$, whereas $\Omega_{G_3}$ 
decreases as $\propto a^{-3}$. 
Hence the contribution of cubic Galileons to $\Omega_{\rm DE}$ 
becomes negligibly small in the late matter era.

The stability of point (a) is known by linearly perturbing 
Eqs.~(\ref{auto1})-(\ref{auto4}) with homogenous perturbations $\delta x$, $\delta \Omega_V$, $\delta \Omega_{G_3}$, and $\delta \Omega_r$ \cite{CLW,CST}. 
The eigenvalues of Jacobian matrix associated with these perturbations are given by 
$-1$, $-(3-2Q^2)/(2-4Q^2)$, 
$-(3+2Q^2)/(1-2Q^2)$, and 
$(3-2Q\lambda-6Q^2)/(1-2Q^2)$. 
The first three eigenvalues are negative for $\lambda$ and $Q$ 
in the ranges (\ref{lamra}) and (\ref{Qra}), 
while the last one is positive. 
Hence the $\phi$MDE corresponds to a saddle point.
This shows that, as long as $\Omega_K$ catches up with
$\Omega_{G_3}$ by the end of radiation era, the solutions 
temporally approach the $\phi$MDE with 
$\Omega_{G_3} \ll  \Omega_K \simeq {\rm constant}$.

There are other kinetically driven fixed points 
characterized by $(x,\Omega_V,\Omega_{G_3},\Omega_r)
=(1/(\sqrt{6}Q \pm 1),0,0,0)$. 
Since $\Omega_m=0$, this point cannot 
be responsible for the matter era. 
The scaling fixed point 
$(x,\Omega_V,\Omega_{G_3},\Omega_r)
=(-\sqrt{6}/(2\lambda), (3-2Q \lambda-6Q^2)/(2\lambda^2),0,0)$ 
is also present, but $\Omega_{\rm DE}=(3-7Q \lambda-12Q^2)/\lambda^2$ 
is larger than the order 1 under the conditions 
(\ref{lamra}) and (\ref{Qra}). 
Hence this scaling solution is irrelevant to the 
matter-dominated epoch. 
This is also the case for the radiation scaling solution 
$(x,\Omega_V,\Omega_{G_3},\Omega_r)
=(-2\sqrt{6}/(3\lambda), 4/(3\lambda^2),0,
1-4(1-2Q \lambda-4Q^2)/\lambda^2)$, 
where $\Omega_{\rm DE}=
4(1-2Q \lambda-4Q^2)/\lambda^2$ exceeds the order 1.  

The fixed point relevant to the dark energy 
domination is given by 
\ba
\hspace{-0.95cm}
& &
({\rm b})~\left( x, \Omega_{V}, \Omega_{G_3}, \Omega_{r} 
\right) \nonumber \\
\hspace{-0.95cm}
& &
=\left( \frac{-\sqrt{6}(\lambda+4Q)}{6(1-Q\lambda-4Q^2)},
\frac{6-\lambda^2-8Q (\lambda+2Q)}{6(1-Q\lambda-4Q^2)^2},0,0 \right),
\label{bpoint}
\ea
with 
\be
\Omega_m=0\,,\quad 
w_{\rm eff}=w_{\rm DE}=
-1+\frac{\lambda^2+6Q \lambda+8Q^2}
{3(1-Q \lambda-4Q^2)},
\label{weffc}
\ee
and $\Omega_{\rm DE}=1$. 
On this fixed point, the quantity $\alpha_{\rM}$ yields 
\be
\alpha_{\rM}^{(\rm b)}=\frac{2Q (\lambda+4Q)}
{1-Q \lambda -4Q^2}\,.
\ee

The point (b) can drive the cosmic acceleration for 
$w_{\rm eff}<-1/3$, which translates to
\be
\lambda^2<2(1-4Q \lambda-8Q^2)\,.
\label{lamcon}
\ee
Under this bound, the four eigenvalues of 
Jacobian matrix of homogeneous perturbations around 
point (b) are all negative. 
Then, after the $\phi$MDE, the solutions finally 
approach the stable point (b) with cosmic acceleration.  
On using the values of $x$ and $\Omega_{V}$ in 
Eq.~(\ref{bpoint}), Eq.~(\ref{auto3}) reduces to
\be
\Omega_{G_3}'=-p\,\Omega_{G_3}\,,\qquad 
p=\frac{(\lambda+4Q)^2}{1-Q \lambda -4Q^2}\,.
\ee
The Galileon density parameter decreases as 
$\Omega_{G_3} \propto a^{-p}$ around point (b).

In the following, we focus on the couplings satisfying
\be
\lambda>0\,,\qquad Q>0\,.
\label{lQcon}
\ee
During the $\phi$MDE, we showed that $\alpha_{\rM}>0$ 
for $Q^2<1/2$. 
Provided $x$ does not change the sign during the 
cosmological evolution from the radiation era to fixed 
point (b), the parameter $\alpha_{\rM}$ is in the range
\be
\alpha_{\rM}=-2\sqrt{6} Qx>0\,,
\label{aMre}
\ee
and hence $x<0$.
The negative value of $x$ is consistent with the fact 
that $\dot{\phi}<0$ when the scalar field rolls down the potential 
with $\lambda>0$. Alternatively, we can consider negative 
values of $\lambda$ and $Q$, in which case $x>0$.
Under the condition (\ref{aMre}), we have 
$Q \dot{\phi}<0$ for $H>0$ and 
hence the quantity $Q \phi$ decreases in time. 
This means that the field $\phi$
satisfies the inequality $Q (\phi-\phi_0)>0$ in the past.
Then, irrespective of the sign of $Q$, the quantity 
$F=e^{-2Q (\phi-\phi_0)/M_{\rm pl}}$ is 
smaller than 1 during the past cosmic expansion history.

\begin{figure}[h]
\begin{center}
\includegraphics[height=3.4in,width=3.4in]{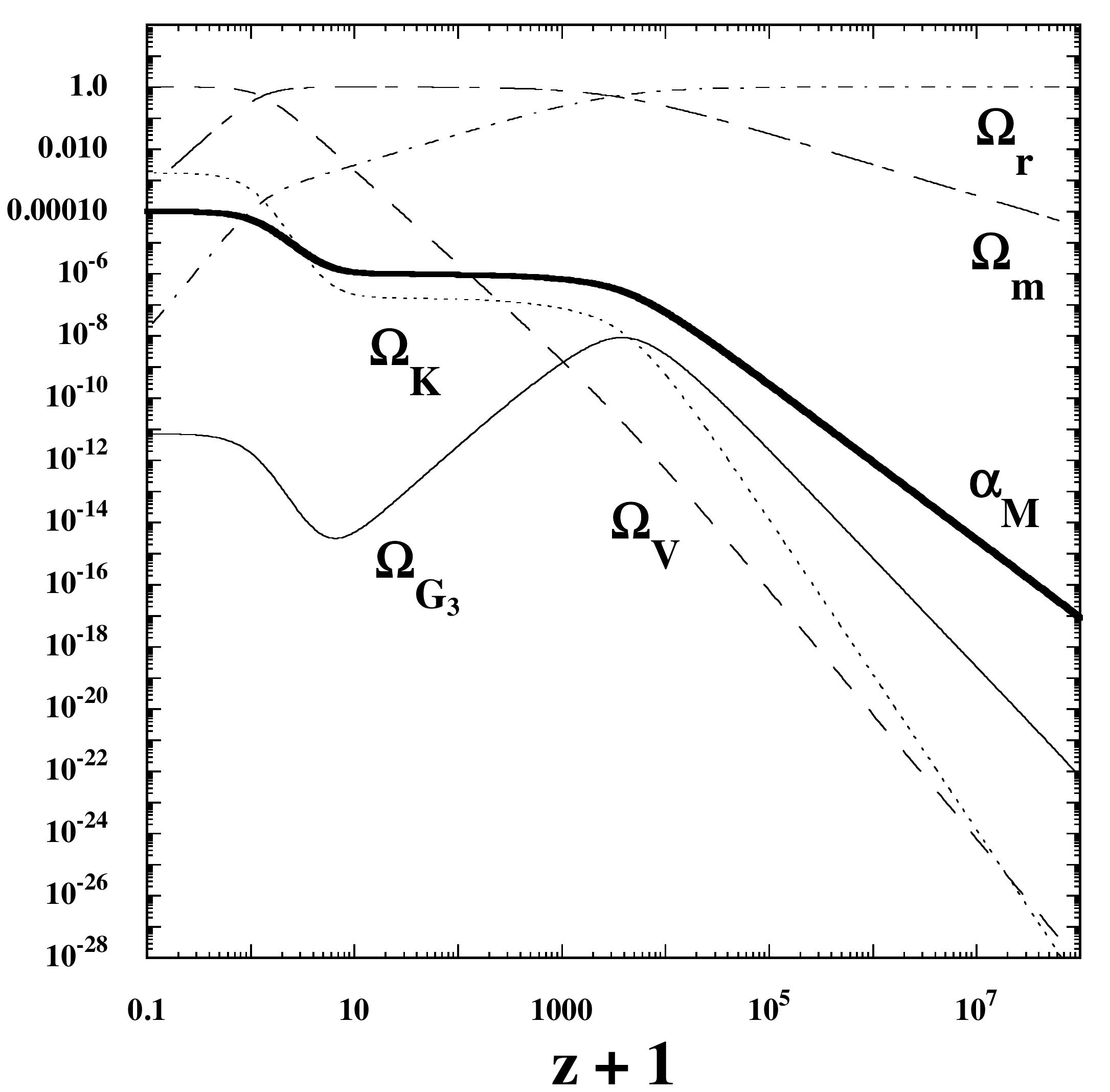}
\end{center}
\caption{\label{fig1}
Evolution of $\Omega_K$, $\Omega_V$, $\Omega_{G_3}$, 
$\Omega_m$, $\Omega_r$, and $\alpha_{\rM}$ versus 
$z+1$ for $Q=5.0 \times 10^{-4}$ and 
$\lambda=0.1$ with the initial conditions 
$x=-1.0 \times 10^{-15}$, $\Omega_V=1.0 \times 10^{-29}$, 
$\Omega_{G_3}=1.0 \times 10^{-23}$, 
$\Omega_r=0.99998$, and $y=1.0$ at the redshift 
$z=1.62 \times 10^8$. The present epoch ($z=0$) is 
identified by the condition $\Omega_{\rm DE}=0.68$.
}
\end{figure}

\begin{figure}[h]
\begin{center}
\includegraphics[height=3.4in,width=3.4in]{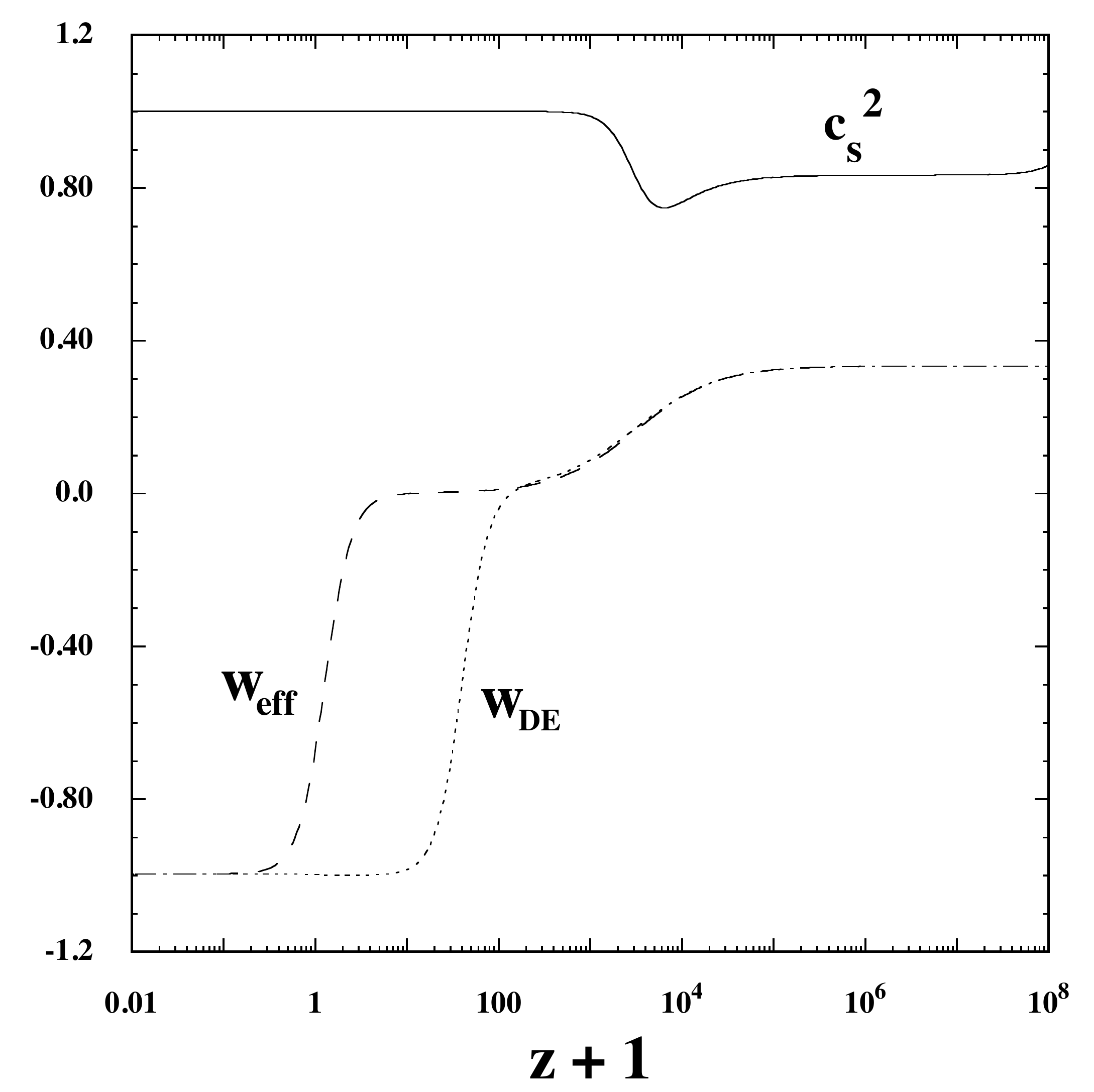}
\end{center}
\caption{\label{fig2}
Evolution of $w_{\rm DE}$, $w_{\rm eff}$, and 
$c_s^2$ versus $z+1$ for the same model parameters 
and initial conditions as those given in the caption 
of Fig.~\ref{fig1}.
}
\end{figure}

In Fig.~\ref{fig1}, we exemplify the evolution of 
$\Omega_K$, $\Omega_V$, $\Omega_{G_3}$, 
$\Omega_r$, $\Omega_m$, and $\alpha_{\rM}$ 
versus $z+1~(=a(t_0)/a(t))$
for $Q=5.0 \times 10^{-4}$ and $\lambda=0.1$.
In this case, the initial value of $\epsilon_{\alpha}$ 
in Eq.~(\ref{hepes}) is $\epsilon_{\alpha}^{(i)}=1.22$, 
so $\epsilon_{\phi}$ starts from the value around $0.72$. 
As estimated from Eq.~(\ref{epphies}), $\epsilon_{\phi}$ 
soon approaches the value $1/2$ during the radiation era.
In Fig.~\ref{fig1}, we can confirm that the evolution of 
$\Omega_K$, $\Omega_V$, $\Omega_{G_3}$, 
$\alpha_{\rM}$ around 
the redshift $10^4 \lesssim z \lesssim 10^8$ 
is approximately given by Eq.~(\ref{Omeevo}). 
In Fig.~\ref{fig2}, we plot the evolution of $w_{\rm DE}$ 
and $w_{\rm eff}$ for the same model parameters and 
initial conditions as those used in Fig.~\ref{fig1}. 
As the analytic estimation (\ref{wdera}) shows, both 
$w_{\rm DE}$ and $w_{\rm eff}$ are close to $1/3$ during the 
deep radiation-dominated epoch. 

In the numerical simulation of Fig.~\ref{fig1}, $\Omega_K$
catches up with $\Omega_{G_3}$ around the 
redshift $z=4.6 \times 10^3$.
Then, the solutions approach the $\phi$MDE 
with the constant kinetic density parameter 
$\Omega_K=2Q^2/[3(1-2Q^2)^2] \simeq 1.7 \times 10^{-7}$ 
with $\alpha_{\rM}=6(1-2Q^2)\Omega_K \simeq 
1.0 \times 10^{-6}$.
As we estimated in Eq.~(\ref{OmeVG}), $\Omega_V$ 
increases during the $\phi$MDE, while $\Omega_{G_3}$ 
decreases. In Fig.~\ref{fig1}, we observe that 
$\Omega_V$ exceeds $\alpha_{\rM}$ around the redshift 
$z=130$. After this moment, $\Omega_V$ becomes 
the dominant contribution to $\Omega_{\rm DE}$. 
As long as $\Omega_V \ll 1$, the terms containing 
$\Omega_V$ in Eqs.~(\ref{hD}) and 
(\ref{epD}) hardly modify the values of $h$ and 
$\epsilon_{\phi}$ during the $\phi$MDE.
In Fig.~\ref{fig1}, we find that the $\phi$MDE with 
nearly constant $\Omega_K$ continues up to the 
redshift $z \approx 10$.

The dark energy equation of state is more sensitive to the dominance 
of $\Omega_V$ over other field density parameters. 
In the regime where the condition 
$\Omega_V \gg \{ \alpha_{\rM}, \Omega_K, 
\Omega_{G_3} \}$ is satisfied, Eq.~(\ref{wdeD}) approximately 
reduces to 
\be
w_{\rm DE} \simeq -1-\frac{2h}{3} 
\frac{1-F}{1-F+\Omega_V F}\,.
\label{wdetra}
\ee
Provided the inequality $\Omega_V F \ll 1-F$ holds 
during the early stage of matter era, it follows that 
$w_{\rm DE} \simeq w_{\rm eff}
=-1-2h/3 \simeq 4Q^2/[3(1-2Q^2)]$. 
After $\Omega_V F$ grows to be larger than $1-F$, 
$w_{\rm DE}$ starts to approach $-1$. 
In Fig.~\ref{fig2}, we can confirm that $w_{\rm DE}$ 
deviates from $w_{\rm eff}$ around the 
same moment at which $\Omega_V$ becomes 
the dominant contribution to $\Omega_{\rm DE}$ 
and that $w_{\rm DE}$ temporally approaches the value close to $-1$.

After the Universe enters the stage of cosmic acceleration, 
the solutions finally reach the fixed point (b). 
For $Q=5.0 \times 10^{-4}$ and $\lambda=0.1$, 
the analytic estimation (\ref{bpoint}) gives the values 
$x=-0.04164$, $\Omega_V=0.9984$, and 
$w_{\rm DE}=w_{\rm eff}=-0.9966$, 
which are in good agreement with the 
numerical results of Figs.~\ref{fig1} and \ref{fig2}. 
In this case, the future asymptotic value of 
$\alpha_{\rM}$ is $1.02 \times 10^{-4}$, 
while its today's value is 
$\alpha_{\rM}(t_0)=5.61 \times 10^{-5}$. 
Taking $h=0.7$ in Eq.~(\ref{aMcon}), this case is within 
the LLR bound of $\alpha_{\rM} (t_0)$.

\begin{figure}[h]
\begin{center}
\includegraphics[height=3.2in,width=3.5in]{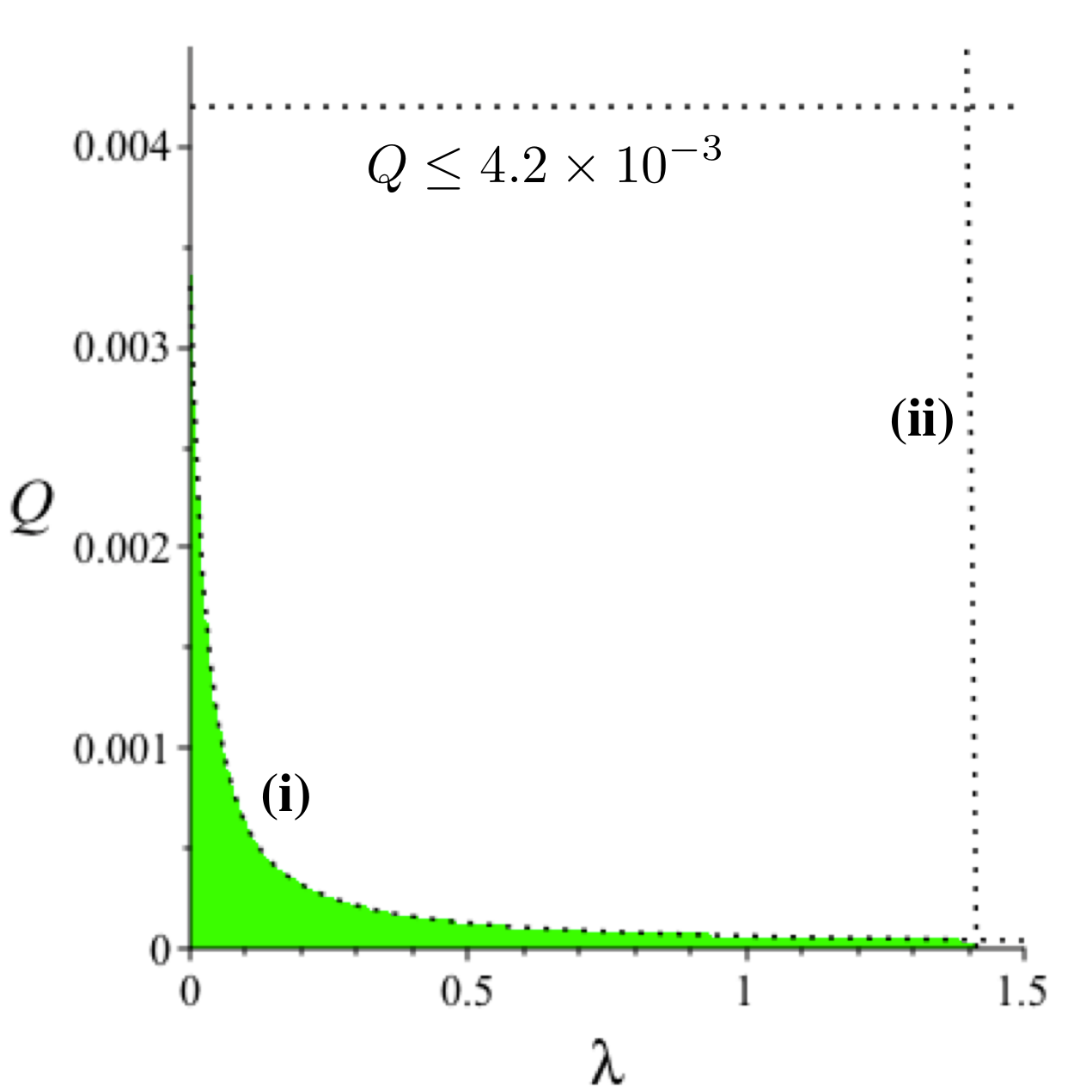}
\end{center}
\caption{\label{fig3}
Parameter space in the $(\lambda, Q)$ plane 
(colored region) consistent with the bound 
(i) $\alpha_{\rM}(t_0) \le 7 \times 10^{-5}$ and 
(ii) the condition for cosmic acceleration of point (b). 
We also show the bound $Q \le 4.2 \times 10^{-3}$ 
arising from the condition 
$\alpha_{\rM}^{({\rm a})} \le 7 \times 10^{-5}$ 
on the $\phi$MDE.}
\end{figure}

{}From Eqs.~(\ref{phiMDE}) and (\ref{bpoint}) we find that 
the inequality $0>x_{(\rm a)}>x_{(\rm b)}$ holds, where 
$x_{(\rm a)}$ and $x_{(\rm b)}$ are the values of 
$x$ on points (a) and (b) respectively.
Then, the quantity $\alpha_{\rM}$ on point (b) is larger than 
that on point (a), such that 
$\alpha_{{\rM}}^{{(\rm b)}}>\alpha_{{\rM}}^{{(\rm a)}}>0$. 
Since $\alpha_{\rM}$ increases from $\alpha_{{\rM}}^{{(\rm a)}}$ 
during the $\phi$MDE to the asymptotic value $\alpha_{{\rM}}^{{(\rm b)}}$ 
in the future, the necessary condition for satisfying 
the LLR bound (\ref{aMcon}) for $h=0.7$ is 
$\alpha_{{\rM}}^{{(\rm a)}} \le 7 \times 10^{-5}$, i.e., 
\be
Q \le 4.2 \times 10^{-3}\,.
\label{Qup}
\ee
Since today's value $\alpha_{{\rM}}(t_0)$ is between 
$\alpha_{{\rM}}^{{(\rm b)}}$ and $\alpha_{{\rM}}^{{(\rm a)}}$, 
the condition (\ref{Qup}) is not sufficient for the compatibility 
with the bound (\ref{aMcon}).

In Fig.~\ref{fig3}, we plot the parameter space 
in the $(\lambda, Q)$ plane constrained from the bound 
$\alpha_{\rM}(t_0) \le 7 \times 10^{-5}$, 
whose border is denoted as the line (i). 
We also depict the region in which the condition 
(\ref{lamcon}) for cosmic acceleration of point (b)
is satisfied, whose border is shown as the line (ii). 
This condition gives the upper limit $\lambda<\sqrt{2}$.
The coupling $Q$ is constrained to be 
\be
Q \le 3.4 \times 10^{-3}\,,
\label{Qup2}
\ee
which is tighter than (\ref{Qup}). 
This significantly improves the upper limit $Q \le 2.6 \times 10^{-2}$ 
following from  the LLR bound $|\alpha_{\rM} (t_0)| \le 0.02$ 
in 2004 \cite{Babi11}.
We note that the bound (\ref{Qup2}) corresponds to the limit 
$\lambda \to 0$. For increasing $\lambda$ from 0, 
the constraint on $Q$ is more stringent than (\ref{Qup2}), e.g., $Q \le 6.2 \times 10^{-4}$ for $\lambda=0.1$ 
and $Q \le 6.3 \times 10^{-5}$ 
for $\lambda=1$. If $\lambda>0.013$, then the recent LLR data give the upper limit of $Q$ tighter than the Cassini 
bound $Q \le 2.4 \times 10^{-3}$ derived for the massless 
scalar field without the Vainshtein screening.

Cosmologically, today's value of $\Omega_{G_3}$ is 
related to the dimensionless coupling $\beta_3$, as
\be
\Omega_{G_3} (t_0)=-6 \sqrt{6} \beta_3\,x(t_0)^3\,.
\ee
The numerical simulation of Fig.~\ref{fig1} corresponds to 
$\Omega_{G_3}(t_0)=1.76 \times 10^{-12}$, 
$x(t_0)=-2.29 \times 10^{-2}$, and 
$\beta_3=9.97 \times 10^{-9}$, with $Q=5.0 \times 10^{-4}$. 
These couplings satisfy the condition (\ref{local}), so the 
Vainshtein mechanism is at work in the solar system.
The existence of $\phi$MDE generally requires that 
$\beta_3 \ll 1$, but still the fifth force can be screened around 
local sources for the product $\beta_3 Q$ 
in the range (\ref{local}).

\begin{figure}[h]
\begin{center}
\includegraphics[height=3.2in,width=3.5in]{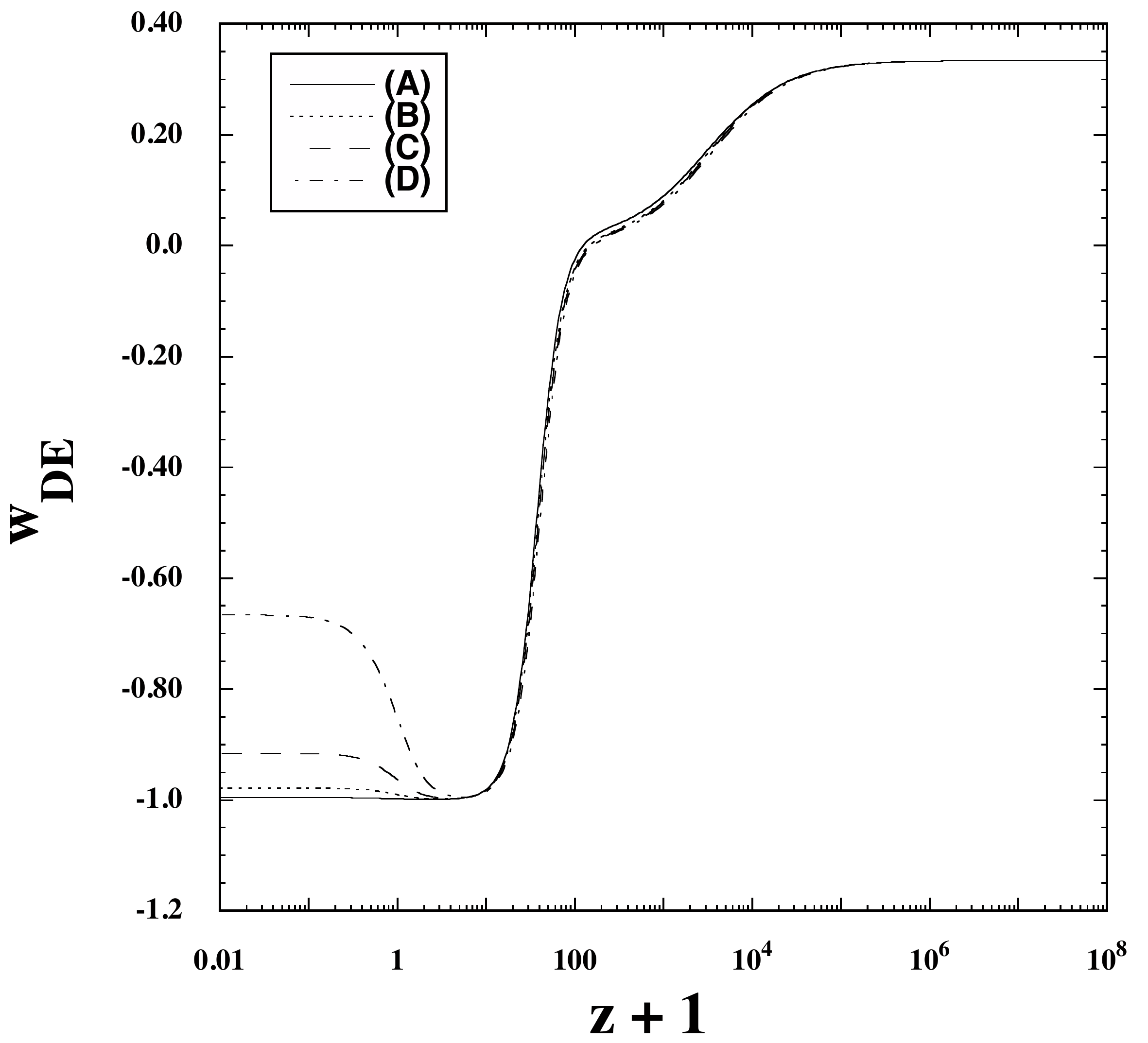}
\end{center}
\caption{\label{fig4}
Evolution of $w_{\rm DE}$ versus $z+1$ for 
(A) $Q=6.20 \times 10^{-4}$, $\lambda=0.1$, 
(B) $Q=2.57 \times 10^{-4}$, $\lambda=0.25$, 
(C)  $Q=1.27 \times 10^{-4}$, $\lambda=0.5$, and 
(D)  $Q=6.32 \times 10^{-5}$, $\lambda=1$. 
The initial conditions of $x$, $\Omega_V$, 
$\Omega_{G_3}$, $\Omega_r$, and $y$ are 
the same as those used in Fig.~\ref{fig1}.}
\end{figure}

In Fig.~\ref{fig4}, we show the evolution of $w_{\rm DE}$ 
for four different combinations of $Q$ and $\lambda$.
In all these cases, $\alpha_{\rM}(t_0)$ is close to the LLR 
upper limit $7 \times 10^{-5}$, with $\beta_3$ 
of order $10^{-8}$.
As we estimated in Eq.~(\ref{wdetra}), $w_{\rm DE}$ 
temporally approaches the value close to $-1$ after 
$\Omega_V$ dominates over other field density 
parameters in the matter era. 
In all the cases plotted in Fig.~\ref{fig4}, 
the minimum values of $w_{\rm DE}$ are close to $-1$.
Even for the case (D), i.e., $\lambda=1$, 
$w_{\rm DE}$ reaches the minimum value 
$-0.9952$ at $z=4.5$. The solutions finally approach 
the fixed point (b), with $w_{\rm DE}$ given by 
Eq.~(\ref{weffc}).
For larger $\lambda$ closer to the border line (ii) 
in Fig.~\ref{fig3}, the deviation of $w_{\rm DE}$ from 
$-1$ at low redshifts is more significant.
This property can be used to distinguish between 
the models with different values of $\lambda$ from 
observations.

Since $\Omega_{G_3}$ and $\Omega_K$ are positive  
with $0<\alpha_{\rM} \ll 1$ from the radiation era to 
the accelerated point (b), the no-ghost condition  
(\ref{Qs}) of scalar perturbations is always satisfied. 
Provided that $1 \gg \Omega_{G_3} \gg \Omega_{K}$ in 
the deep radiation era, the scalar propagation 
speed squared (\ref{cs}) reduces to 
$c_s^2 \simeq (2+\epsilon_{\phi})/3$.
In the numerical simulation of Fig.~\ref{fig2}, the quantity
$\epsilon_{\phi}$ approaches the value $1/2$ around
the redshift $z \approx 10^7$, and hence $c_s^2 \simeq 5/6$ 
for $10^5 \lesssim z \lesssim 10^7$. 
During the late radiation era ($3000 \lesssim z \lesssim 10^5$) 
in which $\Omega_r$ starts to deviate from 1, 
$c_s^2$ temporally decreases due to the decrease 
of $\epsilon_{\phi}$. 
For $\Omega_K \gg \Omega_{G_3}$ we have 
$c_s^2 \simeq 1$ from Eq.~(\ref{cs}). 
Indeed, the approach to this value can be confirmed 
in Fig.~\ref{fig2} 
after the onset of matter era.
Since $c_s^2$ remains positive from the radiation era to 
the asymptotic future, the Laplacian instability of 
scalar perturbations is absent.
We note that the property $c_s^2>0$ also 
holds for the four cases shown in Fig.~\ref{fig4}.

\subsection{Screened cosmology}

We proceed to the case (ii) in which the cubic coupling $\beta_3$ 
is in the range 
\be
|\beta_3| \gg 1\,,
\ee
with positive values of $\lambda$ and $Q$.
As we will see below, the field kinetic energy can be 
suppressed even in the late epoch 
through the cosmological Vainshtein mechanism.

\begin{figure}[h]
\begin{center}
\includegraphics[height=3.4in,width=3.4in]{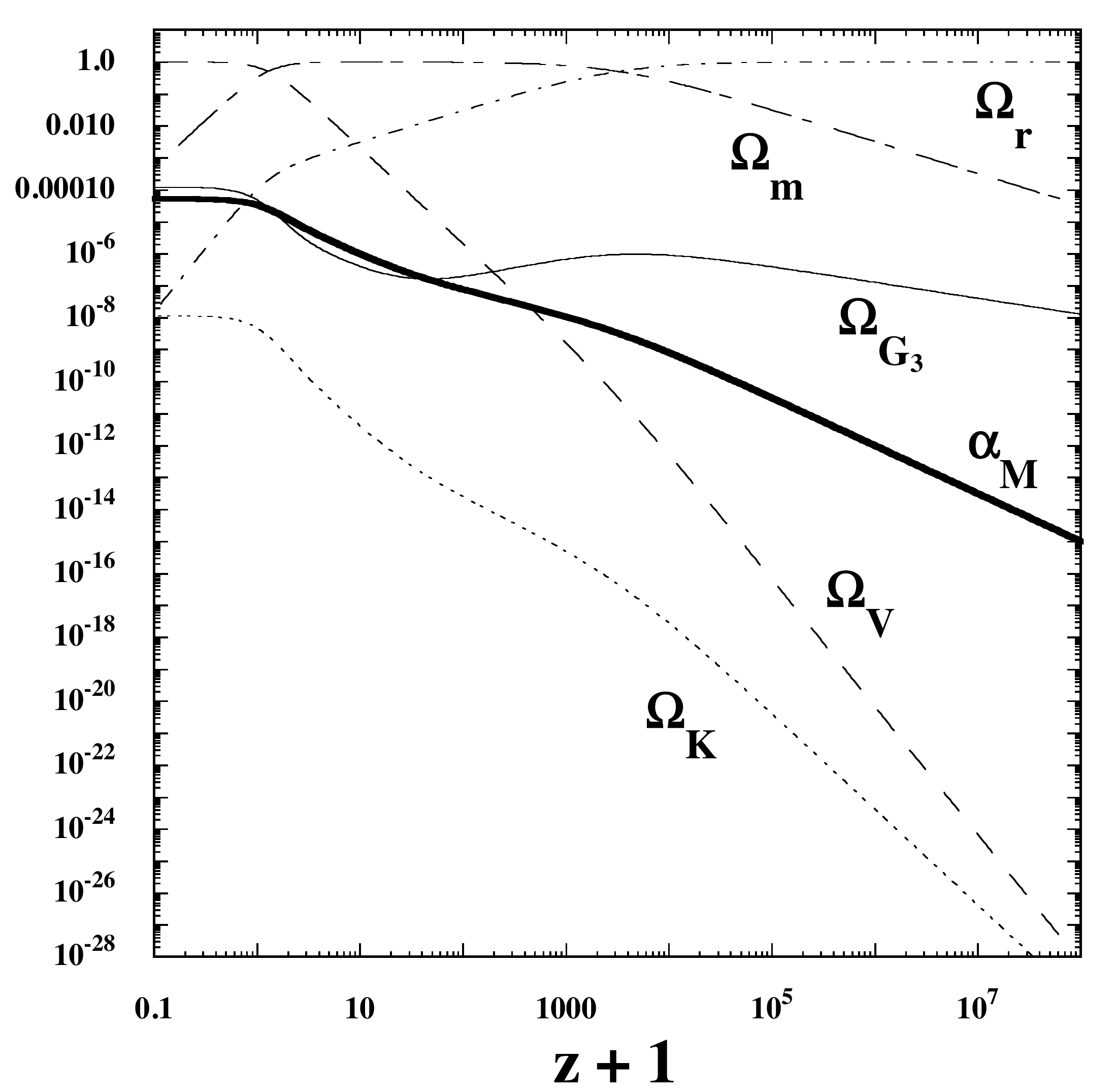}
\end{center}
\caption{\label{fig5}
Evolution of $\Omega_K$, $\Omega_V$, $\Omega_{G_3}$, 
$\Omega_m$, $\Omega_r$, and $\alpha_{\rM}$ versus 
$z+1$ for $Q=0.1$ and $\lambda=1$ with the initial conditions 
$x=-1.0 \times 10^{-15}$, $\Omega_V=1.0 \times 10^{-29}$, 
$\Omega_{G_3}=1.0 \times 10^{-8}$, 
$\Omega_r=0.99998$, and $y=1.0$ at the redshift 
$z=1.62 \times 10^8$. 
}
\end{figure}

\begin{figure}[h]
\begin{center}
\includegraphics[height=3.4in,width=3.4in]{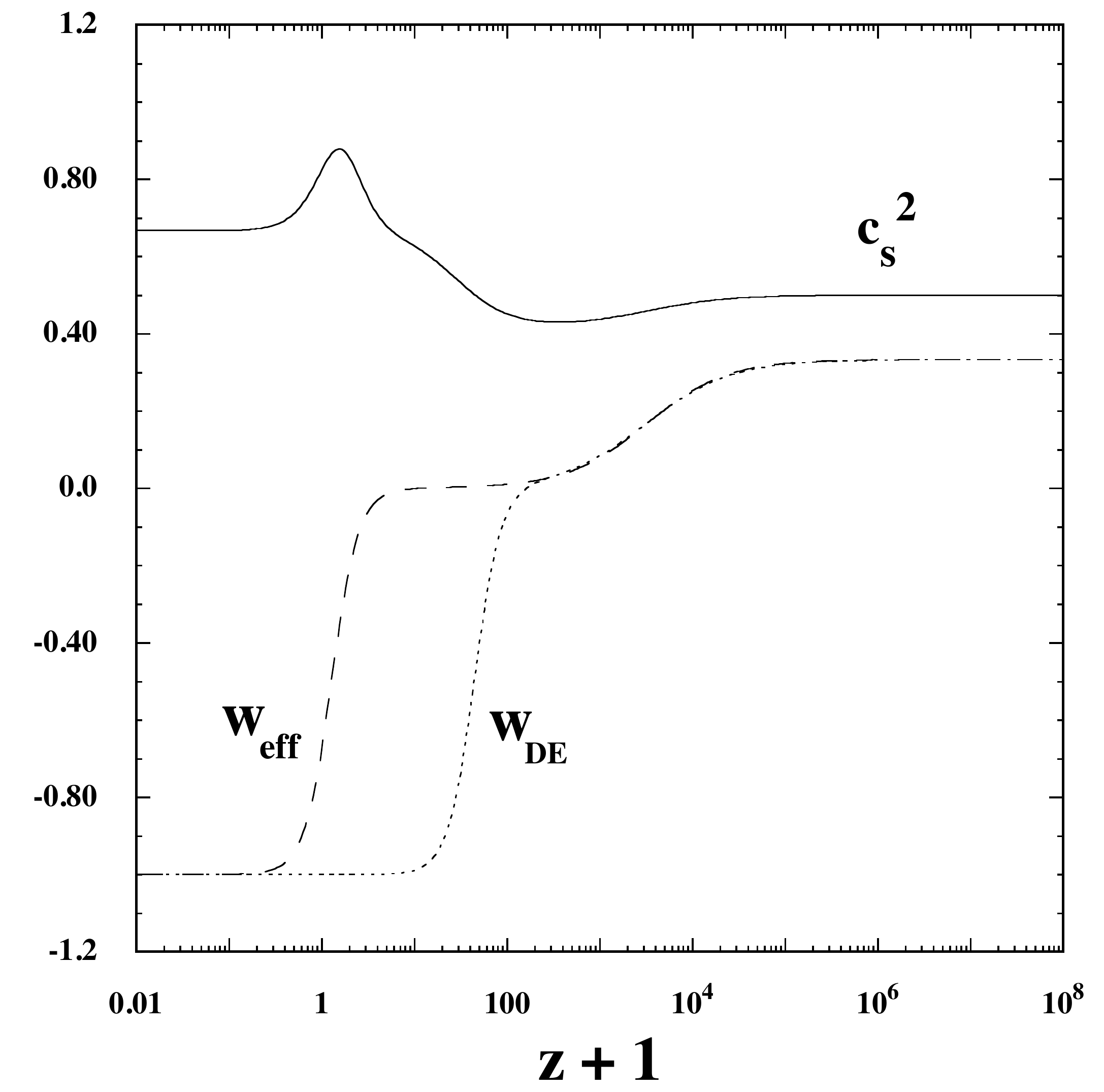}
\end{center}
\caption{\label{fig6}
Evolution of $w_{\rm DE}$, $w_{\rm eff}$, and 
$c_s^2$ versus $z+1$ for the 
same model parameters and initial conditions 
as those used in Fig.~\ref{fig5}.
}
\end{figure}

During the radiation dominance the condition 
(\ref{radcon}) holds, so the quantity $\epsilon_{\phi}$ can be estimated as Eq.~(\ref{hepes}).
The difference from the case discussed in Sec.~\ref{unsec} is 
that $\epsilon_{\alpha}$ is much smaller than 1 due to the largeness 
of $\Omega_{G_3}$ relative to $\alpha_{\rM}$.
Since $\epsilon_{\phi} \simeq -1/2$ during most 
stage of the radiation era, the field density parameters 
and $\alpha_{\rM}$ evolve according to 
Eq.~(\ref{den1}). Indeed, we can confirm this behavior in Fig.~\ref{fig5}, 
where the cubic coupling is $\beta_3=1.0 \times 10^7$. 
Although $\Omega_K$ grows faster than $\Omega_{G_3}$, 
the inequality $\Omega_{G_3} \gg \Omega_K$  holds 
even after the end of radiation era. 
Hence the solutions do not reach the 
$\phi$MDE charactrized by constant $\Omega_K$ 
larger than $\Omega_{G_3}$. 
In Fig.~\ref{fig6}, we observe that both $w_{\rm DE}$ and $w_{\rm eff}$ 
are close to $1/3$ during the radiation dominance.

During the matter-dominated epoch, we study the 
cosmological evolution 
under the conditions:
\ba
& &
\Omega_K \ll \Omega_{G_3} \ll 1\,,\quad 
\alpha_{\rM} \ll 1\,,\quad 
\Omega_V \ll 1\,, \nonumber \\
& &  \Omega_r \ll 1\,,\quad (\lambda/Q)\Omega_V \ll 1\,.
\label{sccon}
\ea
Then,  the quantities defined in Eqs.~(\ref{hD}) and 
(\ref{epD}) reduce to $h \simeq -3/2$ 
and $\epsilon_{\phi} \simeq -3/4+\alpha_{\rM}
/(4\Omega_{G_3})$, respectively.
{}From Eqs.~(\ref{auto1})-(\ref{auto3}), we obtain the 
differential equations for
$\alpha_{\rM}$, $\Omega_V$, and $\Omega_{G_3}$, as 
\ba
\alpha_{\rM}' &\simeq &\frac{\alpha_{\rM}}{4} 
\left( 3+\frac{\alpha_{\rM}}{\Omega_{G_3}} \right)\,,\\
\Omega_{V}' &\simeq & 3\Omega_V\,,\\
\Omega_{G_3}' &\simeq& -\frac{3}{4} 
\left( \Omega_{G_3}-\alpha_{\rM} \right)\,.
\ea
This means that, provided $x<0$, $\alpha_{\rM}$ 
increases during the matter era. 
The density parameter associated with the field potential 
also grows as $\Omega_{V} \propto a^3$. 
On the other hand, $\Omega_{G_3}$ decreases for 
$\Omega_{G_3}>\alpha_{\rM}$, whereas it increases 
for $\Omega_{G_3}<\alpha_{\rM}$. 
In the numerical simulation of Fig.~\ref{fig5}, 
$\Omega_{G_3}$ is larger than $\alpha_{\rM}$ 
at the onset of matter era and hence $\Omega_{G_3}$ decreases 
by the moment at which $\alpha_{\rM}$ catches 
up with $\Omega_{G_3}$. 
After this catch up, $\Omega_{G_3}$ starts to grow. 
The field kinetic density parameter increases as 
$\Omega_K \propto \alpha_{\rM}^2$, but still 
$\Omega_K$ is smaller than $\Omega_{G_3}$ 
around the end of matter era.

In Fig.~\ref{fig5}, we find that $\Omega_V$ dominates over $\Omega_{G_3}$, 
$\Omega_K$, and $\alpha_{\rM}$ for the redshift $z \lesssim 200$.
Then, the dark energy equation of state after the dominance of $\Omega_V$ 
is given by Eq.~(\ref{wdetra}). The numerical simulation of Fig.~\ref{fig6} shows 
that $w_{\rm DE}$ starts to deviate from $w_{\rm eff} \simeq 0$ 
around $z=200$ and then $w_{\rm DE}$ approaches the value close to $-1$ 
for $z \lesssim 10$. 
{}From the radiation dominance to the deep matter era, 
we have
$\epsilon_{\phi} \simeq [\Omega_r-3+(1-\Omega_r)(\alpha_{\rM}/\Omega_{G_3})]/4$ 
under the condition (\ref{sccon}).
Then, the sound speed squared $c_s^2 \simeq (2+\epsilon_{\phi})/3$ can be 
estimated as
\be
c_s^2 \simeq \frac{1}{12} \left[ 5+\Omega_r+\frac{\alpha_{\rM}}{\Omega_{G_3}}
(1-\Omega_r) \right]\,,
\label{csfd}
\ee
which is valid for $z \gg 10$. As $\Omega_r$ starts to deviate from 1 
in the late radiation era, 
$c_s^2$ decreases from the initial value close to $1/2$. 
Since the ratio $\alpha_{\rM}/\Omega_{G_3}$ grows 
in the deep matter era, the term $(\alpha_{\rM}/\Omega_{G_3})(1-\Omega_r)$ in 
Eq.~(\ref{csfd}) starts to increase the value of $c_s^2$. 
Indeed, in the numerical simulation of Fig.~\ref{fig6}, $c_s^2$ 
reaches the minimum value $0.430$ around $z=365$.

In Fig.~\ref{fig5}, we observe that $\Omega_V$, 
$\Omega_{G_3}$, and $\Omega_K$ asymptotically 
approach constants with $\Omega_V ={\cal O}(1) \gg \Omega_{G_3} \gg \Omega_K$. 
In the regime where 
$\Omega_V$ dominates over $\Omega_{G_3}$, 
$\Omega_K$, and $\Omega_r$, 
Eq.~(\ref{auto1}) approximately reduces to  
\be
x' \simeq \frac{x}{4} \left[ 3(1-3\Omega_V) 
-\frac{2\sqrt{6} x}{\Omega_{G_3}} \{ Q+(3Q+\lambda)\Omega_V \} 
\right]\,.
\label{dxeq}
\ee
Then, the solutions approaching a nonvanishing constant  $x$ is given by 
\be
x \simeq -\frac{\sqrt{6} (3\Omega_V-1)}
{4[Q+(3Q+\lambda)\Omega_V]} \Omega_{G_3}\,.
\label{xest}
\ee
Substituting this relation into Eqs.~(\ref{auto2}) and 
(\ref{auto3}), it follows that 
\ba
\Omega_V' &\simeq& 3\Omega_V 
\left( 1-\Omega_V \right)\,,\label{OmeVe} \\
\Omega_{G_3}' &\simeq& 
3 \left( \Omega_V-1 \right)\Omega_{G_3}\,,
\label{OmeG3es}
\ea
which can be integrated to give
\ba
\Omega_V &\simeq& 
\left( 1+c_1 a^{-3} \right)^{-1}\,, \label{Ome1late} \\
\Omega_{G_3} &\simeq& 
c_2 \left( 1+c_1 a^{-3} \right)\,,
\label{OmeG3late}
\ea
where $c_1$ and $c_2$ are constants.
These solutions are valid only at the very late 
cosmological epoch in which $x$ starts to approach a constant.
{}From Eqs.~(\ref{Ome1late}) and (\ref{OmeG3late}), $\Omega_V$ and 
$\Omega_{G_3}$ approach 
the values 1 and $c_2$, respectively. 
Taking the limit $\Omega_{V} \to 1$ in Eq.~(\ref{xest}), 
we can estimate the asymptotic values of $\alpha_{\rM}$ 
and the ratio $\Omega_K/\Omega_{G_3}$, as 
\ba
\alpha_{\rM} &=& \frac{6Q}
{4Q+\lambda} \Omega_{G_3}\,,\label{aMas}\\
\frac{\Omega_K}{\Omega_{G_3}} &=& 
\frac{3}{2(4Q+\lambda)^2} \Omega_{G_3}\,.
\label{aMas2}
\ea
They are in good agreement with the numerical values in 
Fig.~\ref{fig5}, i.e., $\alpha_{\rM}=5.24 \times 10^{-5}$ and  
$\Omega_K/\Omega_{G_3}=9.37 \times 10^{-5}$ with 
$\Omega_{G_3}=1.22 \times 10^{-4}$, so the condition 
$\Omega_{G_3} \gg \Omega_K$ is satisfied. 
We note that, for the other solution $x=0$ in Eq.~(\ref{dxeq}), 
$\Omega_{G_3}$ approaches 0, so this does not lead to the solution 
with $\Omega_{G_3} \gg \Omega_K>0$.

In the numerical simulation of Fig.~\ref{fig5}, today's value of 
$\alpha_{\rM}$ is $3.38 \times 10^{-5}$ and hence this case 
is within the LLR bound (\ref{aMcon}). 
On using Eq.~(\ref{aMas}), the criterion for consistency 
with the LLR experiment is that the asymptotic value of 
$\Omega_{G_3}$ is in the range, 
\be
\frac{6Q}{4Q+\lambda}
\Omega_{G_3} \le 7 \times 10^{-5}\,.
\ee
This is a sufficient condition, so the actual upper bound on 
$\Omega_{G_3}$ is slightly tighter. 
Unlike the case discussed in Sec.~\ref{unsec}, 
the coupling $Q$ is not particularly bounded from above.
Indeed, the numerical simulation of Fig.~\ref{fig5} 
corresponds to $Q=0.1$, but the LLR bound is satisfied. 
This property comes from the fact that the 
cubic Galileon term suppresses the field kinetic energy 
through the cosmological Vainshtein screening, so that the variable 
$x$ in $\alpha_{\rM}=-2\sqrt{6}Qx$ is restricted to be small.
We note that, even though $\Omega_K \ll \Omega_{G_3}$, 
$\Omega_{G_3}$ is much smaller than $\Omega_V$, so 
the cubic Galileon is sub-dominant as the dark energy density. 

The asymptotic value of $\epsilon_{\phi}$ in the future 
is close to $h~(\simeq 0)$ to realize $x'=0$ with $x \neq 0$ in Eq.~(\ref{auto1}). 
Then, the scalar propagation speed squared 
should approach the value 
$c_s^2 \simeq (2+\epsilon_{\phi})/3 \simeq 2/3$, 
which is indeed the case for the numerical simulation 
in Fig.~\ref{fig6}. Since the condition $c_s^2>0$ 
is satisfied from the radiation dominance to the future, 
there is no Laplacian instability of scalar perturbations. 

The numerical simulation of Fig.~\ref{fig6} corresponds to $\lambda=1$, but 
$w_{\rm DE}$ is very close to $-1$ even in the asymptotic 
future. This behavior is different from the case (D) in 
Fig.~\ref{fig4} where the solutions finally 
reach the fixed point (b) with the large deviation 
of $w_{\rm DE}$ from $-1$. 
In the screened cosmology discussed in this section, 
the future asymptotic solution is characterized by 
Eqs.~(\ref{aMas}) and (\ref{aMas2}) with the strongly 
suppressed kinetic energy ($\Omega_K \ll 
\Omega_{G_3} \ll \Omega_V \simeq 1$). 
In this case, the dark energy equation of state is 
given by Eq.~(\ref{wdetra}) with $h \simeq 0$ 
in the asymptotic future and hence 
$w_{\rm DE} \simeq -1$.

\begin{figure}[h]
\begin{center}
\includegraphics[height=3.2in,width=3.5in]{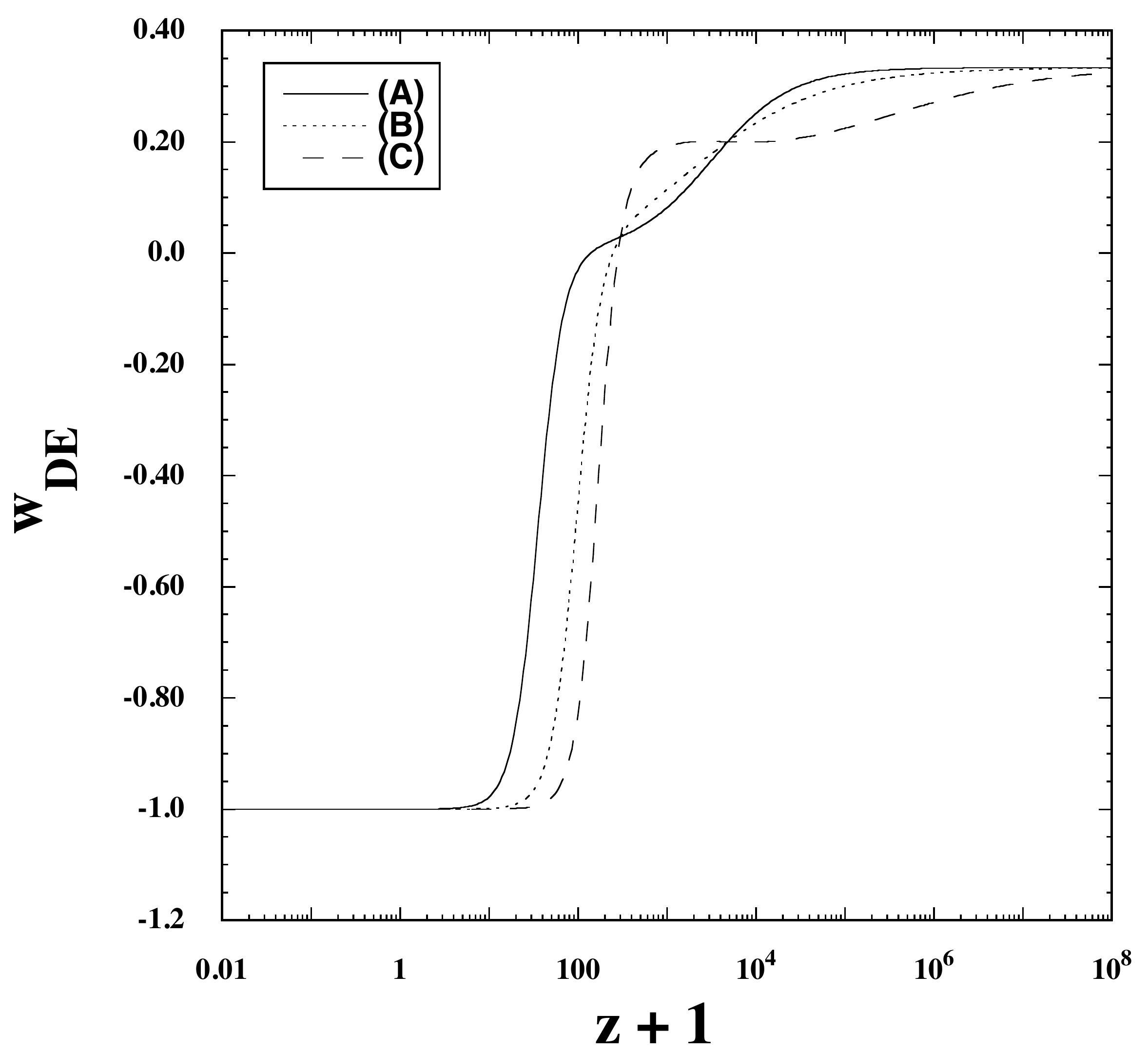}
\end{center}
\caption{\label{fig7}
Evolution of $w_{\rm DE}$ versus $z+1$ for $\lambda=2$ 
for the same initial conditions of $x$, $\Omega_V$, 
$\Omega_{G_3}$, $\Omega_r$, and $y$ as those 
used in Fig.~\ref{fig5}.
Each case correspond to (A) $Q=0.153$, 
(B) $Q=0.010$, and (C) $Q=0.001$.}
\end{figure}

Since the cosmological Vainshtein screening for the 
field kinetic energy efficiently works for $\beta_3 \gg 1$, 
it is possible to realize $w_{\rm DE}$ close to $-1$ 
at low redshifts even for $\lambda>\sqrt{2}$. 
In Fig.~\ref{fig7}, we plot the evolution of $w_{\rm DE}$ for 
$\lambda=2$ with three different values of $Q$, all of which 
correspond to $\beta_3 \simeq 1.0 \times 10^7$. 
Even with $\lambda$ larger than $\sqrt{2}$, $w_{\rm DE}$ is
very close to $-1$ from the redshift $z \approx {\cal O}(10)$ toward the asymptotic future. 
For decreasing $Q$, the deviation of 
$F=e^{-2Q (\phi-\phi_0)/M_{\rm pl}}$ from 1 
tends to be smaller in the past and hence the solutions 
enter the regime $\Omega_V F> 1-F$ at earlier time.
Then, from Eq.~(\ref{wdetra}), the approach of $w_{\rm DE}$ 
to $-1$ occurs at higher redshifts. 
In case (A) of Fig.~\ref{fig7} we have 
$\alpha_{\rM}(t_0)=6.98 \times 10^{-5}$, so this is 
close to the LLR upper limit (\ref{aMcon}). 
For decreasing $Q$ with given values of $\beta_3$ and $\lambda$, 
$\alpha_{\rM}(t_0)$ gets smaller, 
e.g., $\alpha_{\rM}(t_0)=3.87 \times 10^{-6}$ and $\alpha_{\rM}(t_0)=3.82 \times 10^{-7}$ in cases (B) and (C) of Fig.~\ref{fig7}, respectively. 
For smaller $\alpha_{\rM}(t_0)$, the models mimic the 
$\Lambda$CDM behavior ($w_{\rm DE}=-1$) from earlier 
cosmological epochs to today.

\section{Modified gravitational wave propagation}
\label{GWsec}

In this section, we study the modified GW propagation induced 
by the nonminimal coupling $F(\phi)R$ and compute the difference between 
GW and luminosity distances for the dark energy cosmology 
discussed in Sec.~\ref{numesec}.
The perturbed line element containing tensor perturbations $h_{ij}$ 
on the flat FLRW background is given by 
\be
{\rm d} s^2=-{\rm d}t^2+a^2(t) 
\left( \delta_{ij}+h_{ij} \right) 
{\rm d} x^i {\rm d}x^j\,.
\ee
To satisfy the transverse and traceless conditions 
$\partial^j h_{ij}=0$ and ${h_i}^i=0$, we choose the nonvanishing 
components of $h_{ij}$, as $h_{11}=h_1(t,z)$, $h_{22}=-h_1(t,z)$ 
and $h_{12}=h_{21}=h_2(t,z)$.
Expanding the action (\ref{action}) up to quadratic order 
in $h_{ij}$ and integrating it by parts, the resulting 
second-order action of tensor perturbations yields \cite{Horn2,DT12,KT18}
\be
{\cal S}_t^{(2)}=\int {\rm d}t\,{\rm d}^3x \sum_{i=1}^2
\frac{M_{\rm pl}^2}{4} F(\phi) a^3 
\left[ \dot{h}_i^2- \frac{1}{a^2} (\partial h_i)^2 
\right]\,.
\label{St2}
\ee
In general, the speed $c_t$ of tensor perturbations appears 
as the spatial derivative term 
$-(c_t^2/a^2)(\partial h_i)^2$ in the square bracket 
of Eq.~(\ref{St2}). In our theory $c_t^2$ is equivalent to 1, so 
it automatically satisfies the observational bound of 
GW propagation speed \cite{GW170817}.

In Fourier space with the coming wavenumber $k$, 
the two polarization modes $h_i$ (where $i=1, 2$) 
obey the wave equation,
\be
\ddot{h}_i+H \left( 3+\alpha_{\rM} \right)\dot{h}_i
+\frac{k^2}{a^2}h_i=0\,.
\label{hieq}
\ee
By defining 
\be
\hat{h}_i  \equiv a_{{\rm GW}} h_i\,,\qquad  
a_{\rm GW} \equiv \sqrt{F}a\,,
\ee
Eq.~(\ref{hieq}) can be expressed in the form 
\be
\frac{{\rm d}^2\hat{h}_i}{{\rm d}\eta^2}
+\left( k^2 -\frac{1}{a_{\rm GW}} 
\frac{{\rm d}^2 a_{\rm GW}}{{\rm d} \eta^2}
\right) \hat{h}_i=0\,,
\label{hieq2}
\ee
where $\eta=\int a^{-1} dt$ is the
conformal time.

For the physical wavelength much smaller than the Hubble radius 
($k/a \gg H$), the second term in the parenthesis of 
Eq.~(\ref{hieq2}) can be ignored relative to $k^2$.
Then, the solution to Eq.~(\ref{hieq2}) 
is simply given by a plane wave with a constant 
amplitude ($\hat{h}_i \simeq e^{\pm i k\eta}$).
The amplitude of $h_i=\hat{h}_i/a_{\rm GW}$ decreases 
in proportion to $1/a_{\rm GW}$. 
The GW produced by a binary inspiral (point particles 
with two masses $m_1$ and $m_2$) at redshift 
$z$ with the comoving distance $r$ from 
an observer has the amplitude \cite{Michele}:
\be
h_A(z)=\frac{4}{a(t_s)r} 
\left( \frac{G_{\rm N}(t_s) M_c}{c^2} \right)^{5/3} 
\left( \frac{\pi f_s}{c} \right)^{2/3}\,,
\label{hA}
\ee
where $t_s$ is the time at emission, $G_{\rm N}(t_s)=G/F(t_s)$ 
is the screened gravitational coupling at $t=t_s$
with $G=1/(8 \pi M_{\rm pl}^2)$, 
$M_c=(m_1 m_2)^{3/5}/(m_1+m_2)^{1/5}$ 
is the chirp mass, and $f_s$ is the 
frequency measured by the clock of source. 
We note that the speed of light $c$ is explicitly written 
in Eq.~(\ref{hA}).
Today's GW amplitude $h_A(0)$ observed at time $t_0$ 
is related to $h_A(z)$, as 
$h_A(0)=[a_{\rm GW}(t_s)/a_{\rm GW}(t_0)]h_A(z)$.  
On using the property $a_{\rm GW}(t_0)=a (t_0)$, 
it follows that 
\be
h_A(0)=\frac{a_{\rm GW}(t_s)}{a(t_s)} 
\frac{1}{F(t_s)^{5/3}}
h_{A,{\rm GR}}(0)\,,
\label{hA0}
\ee
where 
\be
h_{A,{\rm GR}} (0)=\frac{4}{a(t_0)r} 
\left( \frac{G M_c}{c^2} \right)^{5/3} 
\left( \frac{\pi f_s}{c} \right)^{2/3}
\label{hGR}
\ee
is the observed GW amplitude in GR.
On the flat FLRW background, the luminosity distance 
from the observer to the source is given by 
$d_L(z)=(1+z)a(t_0)r$.
By using $d_L(z)$ and the observed GW frequency 
$f_{\rm obs}=f_s/(1+z)$, one can write Eq.~(\ref{hGR}) in the form 
\be
h_{A,{\rm GR}} (0)=\frac{4}{d_L(z)} 
\left( \frac{G {\cal M}_c}{c^2} \right)^{5/3} 
\left( \frac{\pi f_{\rm obs}}{c} \right)^{2/3}\,,
\label{hGR2}
\ee
where ${\cal M}_c \equiv (1+z)M_c$. 
Substituting Eq.~(\ref{hGR2}) into Eq.~(\ref{hA0}), 
the observed GW amplitude is expressed as
\be
h_{A} (0)=\frac{4}{d_{\rm GW}(z)} 
\left( \frac{G_{\rm N}(t_s) {\cal M}_c}{c^2} \right)^{5/3} 
\left( \frac{\pi f_{\rm obs}}{c} \right)^{2/3}\,,
\label{hGR3}
\ee
where 
\be
d_{\rm GW}(z)=d_L(z) \frac{a(t_s)}{a_{\rm GW}(t_s)}
=\frac{d_L(z)}{\sqrt{F(t_s)}}\,.
\label{dGWL}
\ee
On using Eq.~(\ref{alM}), the quantity $F$ at redshift 
$z$ is generally expressed as   
\be
F(z)=\exp \left[ -\int_0^z \frac{\alpha_{\rM} (\tilde{z})}
{1+\tilde{z}} {\rm d} \tilde{z} \right]\,.
\ee
Then, the relative ratio between $d_{\rm GW}(z)$ and 
$d_L(z)$ yields 
\be
\frac{d_{\rm GW}(z)}{d_L(z)}=
\exp \left[ \int_0^z \frac{\alpha_{\rM} (\tilde{z})}
{2(1+\tilde{z})} {\rm d} \tilde{z} \right]\,.
\label{dgLra}
\ee
If $\alpha_{\rM}(z)>0$, then $d_{\rm GW}(z)>d_L(z)$ for $z>0$. 
For positive $\alpha_{\rM}(z)$, which is the case for our 
nonminimally coupled dark energy scenario, there is the 
LLR bound $\alpha_{\rM}(0) \le \alpha_{\rm max}$, 
where $\alpha_{\rm max}=7 \times 10^{-5}$.
Provided that the past value of $\alpha_{\rM}(z)$ 
is smaller than $\alpha_{\rM}(0)$, the ratio (\ref{dgLra}) 
is in the range 
\be
\frac{d_{\rm GW}(z)}{d_L(z)} \le 
\left( 1+z \right)^{\alpha_{\rm max}/2}\,.
\label{dgLra2}
\ee
Expanding the term 
$\left( 1+z \right)^{\alpha_{\rm max}/2}$ around 
$\alpha_{\rm max}=0$, it follows that 
\be
\mu_d(z) \equiv
\frac{d_{\rm GW}(z)}{d_L(z)}-1 \lesssim 
\frac{\alpha_{\rm max}}{2}  \ln 
\left( 1+z \right)\,,
\label{dgLra2}
\ee
where we ignored the terms 
higher than the order $\alpha_{\rm max}$.
Substituting $\alpha_{\rm max}=7 \times 10^{-5}$ into 
the right hand side of Eq.~(\ref{dgLra2}), we have 
$(\alpha_{\rm max}/2) \ln \left( 1+z \right)=1.6 \times 10^{-4}$ 
at $z=100$.
Then, the quantity $\mu_d (z)$ is constrained to be
\be
\mu_d(z)
\lesssim 10^{-4}\,,\qquad 
({\rm for}~0<z<100)\,.
\label{dgLra3}
\ee
This is the maximum allowed difference between 
$d_{\rm GW}(z)$ and $d_L(z)$ constrained from 
the LLR data. 

\begin{figure}[h]
\begin{center}
\includegraphics[height=3.2in,width=3.5in]{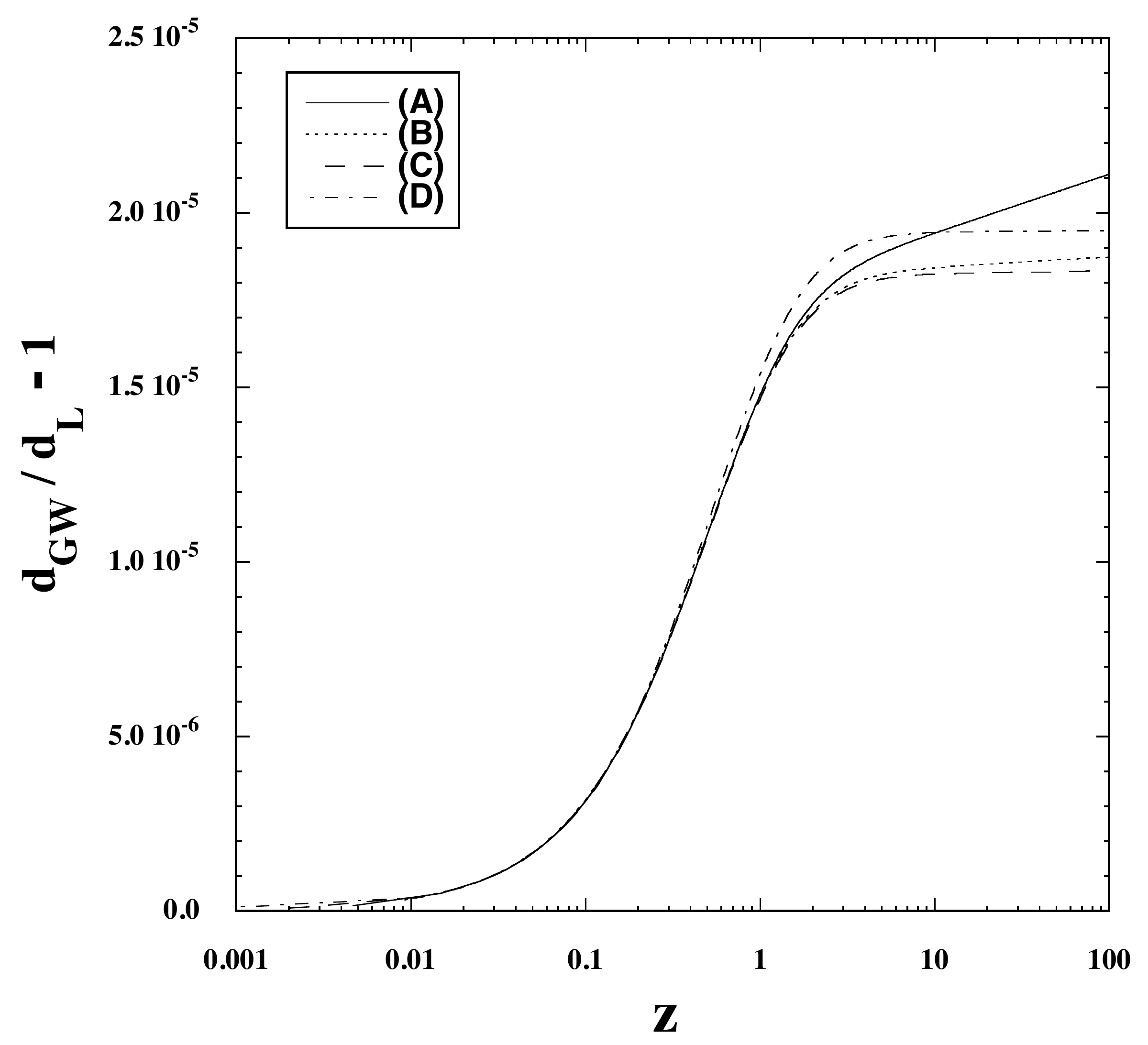}
\end{center}
\caption{\label{fig8}
The relative difference $d_{\rm GW}(z)/d_L(z)-1$ versus $z$ 
corresponding to the cases (A), (B), (C), (D) shown  
in Fig.~\ref{fig4}.}
\end{figure}

\begin{figure}[h]
\begin{center}
\includegraphics[height=3.2in,width=3.5in]{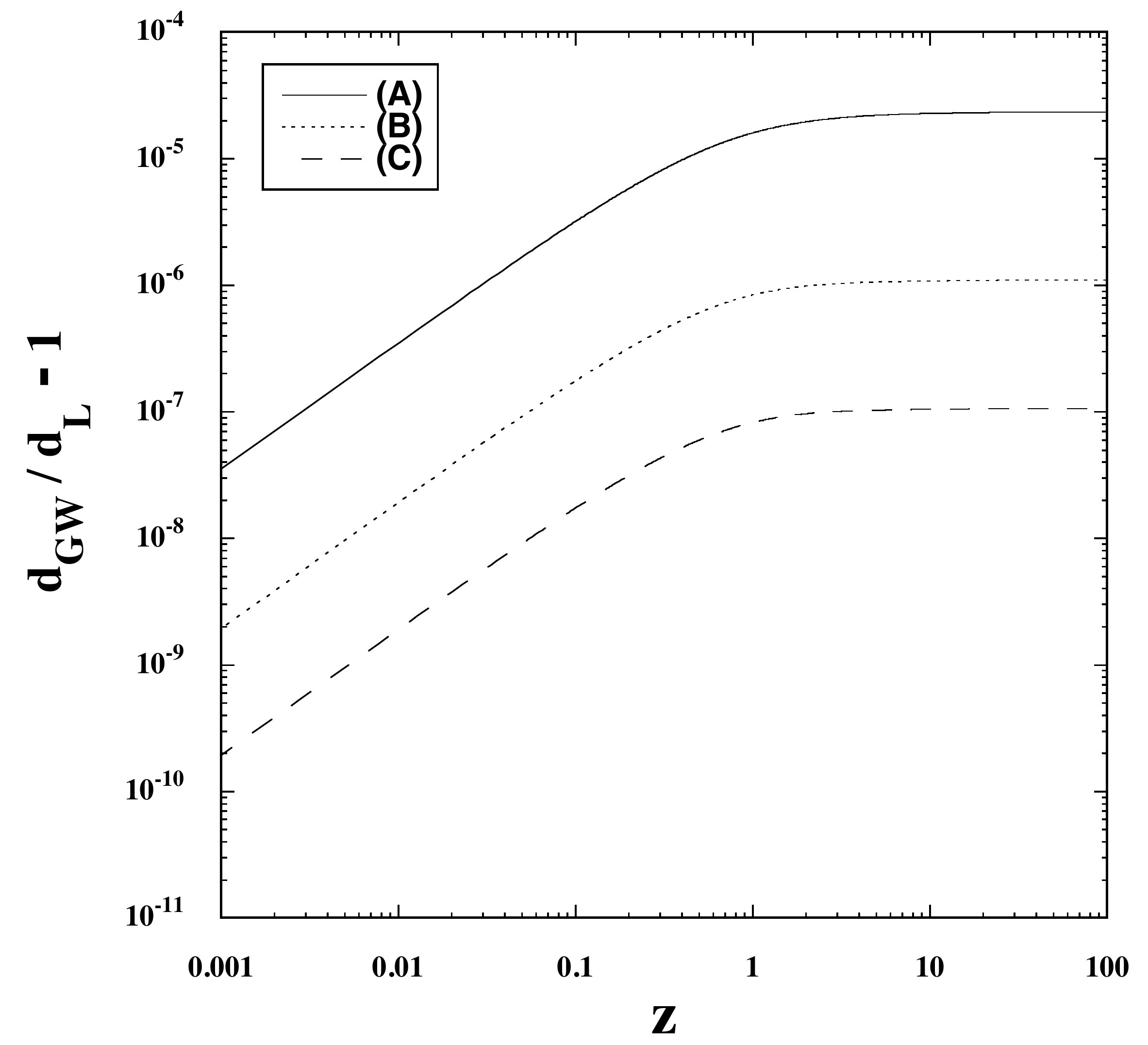}
\end{center}
\caption{\label{fig9}
The relative difference $d_{\rm GW}(z)/d_L(z)-1$ versus $z$ 
corresponding to the cases (A), (B), (C) shown  
in Fig.~\ref{fig7}.}
\end{figure}

For concreteness, let us consider the nonminimally 
coupled dark energy scenario given by the action (\ref{action}).
{}From Eq.~(\ref{dGWL}), we have 
\be
\frac{d_{\rm GW}(z)}{d_L(z)}
=e^{Q[\phi (z)-\phi_0]/M_{\rm pl}}\,.
\label{dra}
\ee
The change of $\phi$ from the redshift $z$ to 
today leads to the difference between 
$d_{\rm GW}(z)$ and $d_L(z)$. 
As we studied in Sec.~\ref{numesec},  there are two 
qualitative different cases: (i) $|\beta_3| \ll 1$ 
with the $\phi$MDE, and (ii) $|\beta_3| \gg 1$ 
without the $\phi$MDE. 

In case (i), the LLR data place the tight upper 
limit (\ref{Qup2}) on the coupling constant $Q$. 
In Fig.~\ref{fig8}, we plot $\mu_d(z)=d_{\rm GW}(z)/d_L(z)-1$ in the redshift range $0<z<100$ for four different combinations of $Q$ 
and $\lambda$. Each plot corresponds to cases (A), (B), (C), (D) 
shown in Fig.~\ref{fig4}. In all these cases, 
the LLR bound is marginally satisfied, i.e., 
$\alpha_{\rM}(0) \simeq 7 \times 10^{-5}$.
For the redshift $z<1$, the values of $\mu_d(z)$ are 
similar to each other between the four cases,  
with  $\mu_d \simeq 1.5 \times 10^{-5}$ at $z=1$.
The difference starts to appear for $z>1$, but the orders 
of $\mu_d(z)$ at $z=100$ are still $10^{-5}$. 
As we estimated in Eq.~(\ref{aMa}), the value of 
$\alpha_{\rM}$ during the $\phi$MDE 
is of order $4Q^2$ and hence 
$\alpha_{\rM}^{({\rm a})} \le 4.6 \times 10^{-5}$ 
under the bound (\ref{Qup2}).
Since $\alpha_{\rM}^{({\rm a})}$ is
smaller than today's value $\alpha_{\rM}(0)$, 
the main contribution to the ratio (\ref{dgLra}) 
comes from $\alpha_{\rM}(z)$ at low redshifts.
Since $\alpha_{\rM}(z)$ at $z \le 1$ is not much 
different from today's value 
$\alpha_{\rM}(0) \simeq 7 \times 10^{-5}$ in 
the numerical simulation of Fig.~\ref{fig8}, 
the maximum value of $\mu_d$ for $z \gg 1$
can be estimated by substituting $z=1$ 
into Eq.~(\ref{dgLra2}), i.e., 
$\mu_d \lesssim {\cal O}(10^{-5})$.
Indeed, this crude estimation is consistent with 
the numerical values of $\mu_d$ at $z \gg 1$ 
in Fig.~\ref{fig8}.
If $\alpha_{\rM}(0)$ is smaller than $7 \times 10^{-5}$, 
the resulting values of $\mu_d$ at high redshifts are 
less than the order $10^{-5}$.

In case (ii), the upper limit of $Q$ is not particularly constrained from the LLR experiment, but the cosmological Vainshtein screening leads to the strong suppression of $\dot{\phi}$. 
The case (A) in Fig.~\ref{fig9}, which corresponds to 
$Q=0.153$ and $\lambda=2$, is marginally within the
LLR bound. In this case, the value of $\alpha_{\rM}$ 
for $z \gg 1$ is of order $10^{-5}$. 
As we see in Fig.~\ref{fig5}, $\alpha_{\rM}$ 
rapidly decreases toward the asymptotic past  
and hence the main contribution to $\mu_d(z)$ again 
comes from $\alpha_{\rM}(z)$ 
at $z \le {\cal O}(1)$.
In cases (B) and (C) of Fig.~\ref{fig9}, which 
correspond to the couplings $Q=0.01$ and $Q=0.001$, 
today's values of $\alpha_{\rM}$ are smaller than
that in case (A) by one and two orders of magnitude, 
respectively.
In cases (B) and (C), the numerical values of $\mu_d(z)$ 
at $z=100$ are $1.1 \times 10^{-6}$ and $1.1 \times 10^{-7}$, 
respectively, so the order difference of $\alpha_{\rM}(0)$ directly affects $\mu_d$ at high redshifts. 

{}From the above discussion, we have $\mu_d(z) \le {\cal O}(10^{-5})$ 
for $0<z<100$ in both unscreened and screened cosmological backgrounds. 
This property is mostly attributed to the fact that 
the value of $\alpha_{\rM}$ at low redshifts is tightly 
limited by the LLR bound.
Unless the ratio $d_{\rm GW}(z)/d_L(z)$ is measured in 
high accuracy, it is challenging to observationally 
distinguish nonminimally coupled theories from minimally coupled theories.
 
\section{Conclusions}
\label{consec}

We studied how the recent LLR measurement constrains
nonminimally coupled dark energy models given by the action (\ref{action}). 
The existence of nonminimal coupling of the form $F(\phi)R$, where 
$F(\phi)=e^{-2Q(\phi-\phi_0)/M_{\rm pl}}$, gives rise to the 
propagation of fifth forces characterized by the coupling constant 
$Q$ with nonrelativistic matter. For a massless scalar field without 
derivative interactions, the coupling is constrained to be 
in the range $|Q| \le 2.4 \times 10^{-3}$ from the Cassini experiment. 
The cubic Galileon coupling $\beta_3 M^{-3} X \square \phi$ 
allows one to recover the Newtonian behavior in over-density regions 
even for $|Q|>2.4 \times 10^{-3}$. 
Since the late-time dominance of Galileons as the dark energy density
generally leads to the incompatibility with observations, 
we considered the potential $V(\phi)$ of a light scalar field.

In local regions of the Universe, the Galileon self-interaction screens 
fifth forces within the Vainshtein radius (\ref{rV}).
The Vainshtein mechanism is at work within the solar system 
for the cubic coupling in the range $|\beta_3 Q| \gg 10^{-17}$.
In spite of the screened scalar-matter interaction, 
the time variation of $\phi$ associated with the dynamics 
of dark energy survives in the expression of gravitational 
coupling $G_{\rm N}$ in over-density regions, with the form 
$G_{\rm N}=1/[8\pi M_{\rm pl}^2F(\phi)]$. 
The recent LLR data placed the tight constraint 
(\ref{Gbou}) on the time variation of $G_{\rm N}$, 
which translates to the bound (\ref{aMcon}) on today's 
value of $\alpha_{\rM}=\dot{F}/(HF)$.

To investigate the evolution of $\alpha_{\rM}$ as well as field density parameters $\Omega_K, \Omega_V, \Omega_{G_3}$, we expressed dynamical equations of motion on the flat FLRW background in the 
autonomous form given by (\ref{auto1})-(\ref{auto4}).
In addition to the dark energy equation of state $w_{\rm DE}$, 
we also considered the quantities $q_s$ and $c_s^2$ to ensure 
the absence of ghosts and Laplacian instabilities.
Together with Eq.~(\ref{dy}), the dynamical 
background equations of motion can be applied to any scalar 
potential $V(\phi)$.

In Sec.~\ref{numesec}, we studied the cosmological dynamics 
in details for the exponential potential (\ref{exp}). 
For the cubic coupling satisfying the condition (\ref{local}), 
$\Omega_{G_3}$ can dominate over $\Omega_K$ in the 
radiation-dominated epoch. 
We showed that, under the conditions 
$|\alpha_{\rM}| \gg \Omega_{G_3}$ and 
$|\alpha_{\rM}| \ll \Omega_{G_3}$, the field density parameters 
and $ |\alpha_{\rM}|$ evolve as Eqs.~(\ref{Omeevo}) and (\ref{den1}), 
respectively, during the radiation era. 
After the onset of matter dominance, there are two qualitatively 
different cases: (i) unscreened cosmology with 
$|\beta_3| \ll 1$, and (ii) screened cosmology 
with $|\beta_3| \gg 1$.

In case (i), there is the kinetically driven $\phi$MDE 
in which $\alpha_{\rM}$ is given by 
$\alpha_{\rM}^{({\rm a})}=4Q^2/(1-2Q^2)$. 
The solutions finally approach the fixed point (b) with 
cosmic acceleration at which $\alpha_{\rM}$ is 
equivalent to $\alpha_{\rM}^{({\rm b})}
=2Q(\lambda+4Q)/(1-Q \lambda-4Q^2)$. 
For positive $\lambda$ and $Q$ the inequality
$\alpha_{\rM}^{({\rm b})}>\alpha_{\rM}^{({\rm a})}>0$ holds, 
so the necessary condition for consistency with 
the LLR bound (\ref{aMcon}) corresponds to 
$\alpha_{\rM}^{({\rm a})} \le 7 \times 10^{-5}$, i.e., 
$Q \le 4.2 \times 10^{-3}$. Applying today's bound 
$\alpha_{\rM}(t_0) \le 7 \times 10^{-5}$ to case (i), 
the coupling is constrained to be 
$Q \le 3.4 \times 10^{-3}$ in the limit $\lambda \to 0$. 
As we see in Fig.~\ref{fig3}, for increasing $\lambda$, 
the upper bound on $Q$ is tighter than the bound 
$Q \le 3.4 \times 10^{-3}$.
We also showed that $w_{\rm DE}$ temporally approaches the 
value close to $-1$ during the matter era after the dominance 
of the term $\Omega_V F$ over $1-F$. 
For larger $\lambda$, the deviation of $w_{\rm DE}$ from 
$-1$ on the attractor point (b) tends to be larger, see 
Fig.~\ref{fig4}.

In case (ii), the cosmological Vainshtein screening of field kinetic energy 
is at work, so the condition $\Omega_K \ll \Omega_{G_3}$ 
is satisfied even after the end of radiation dominance.
As we observe in Fig.~\ref{fig5}, $\alpha_{\rM}$ grows 
during the matter era and finally approaches a constant 
related to $\Omega_{G_3}$, as 
$\alpha_{\rM}=6Q \Omega_{G_3}/(4Q+\lambda)$. 
Provided that this asymptotic value of $\alpha_{\rM}$ is 
smaller than the order $10^{-4}$, the case (ii) can be 
consistent with today's LLR bound (\ref{aMcon}). 
Since $\Omega_{G_3}$ is much smaller than
$\Omega_V$ today, the coupling $Q$ is not 
particularly bounded from above. 
The field kinetic energy is strongly suppressed by 
the cosmological Vainshtein screening, i.e., 
$\Omega_K \ll \Omega_{G_3} \ll \Omega_V$, so 
it is possible to realize $w_{\rm DE}$ very close to $-1$  
at low redshifts even for $\lambda>\sqrt{2}$, see 
Fig.~\ref{fig7}.
This behavior is different from that in case (i) where
$w_{\rm DE}$ deviates from $-1$ in the asymptotic future 
for increasing $\lambda$ in the range $\lambda<\sqrt{2}$.

In Sec.~\ref{GWsec}, we derived the relation between the GW and luminosity distances in the form (\ref{dGWL}). 
In terms of the parameter $\alpha_{\rM}$, the ratio between 
$d_{\rm GW}(z)$ and $d_L(z)$ is given by Eq.~(\ref{dgLra}). 
Provided that $\alpha_{\rM}(z)$ in the past is smaller than 
today's value $\alpha_{\rM}(0)$, the LLR experiment gives the upper limit on 
the relative difference $\mu_d(z)=d_{\rm GW}(z)/d_L(z)-1$ 
as Eq.~(\ref{dgLra2}). We computed the quantity $\mu_d(z)$ 
for the nonminimally coupled dark energy scenario discussed
in Sec.~\ref{numesec} and showed that $\mu_d(z)$ for 
$z \geq {\cal O}(1)$ is mostly determined by today's value 
of $\alpha_{\rM}$. For $\alpha_{\rM}(0)$ 
close to the LLR upper limit $7 \times 10^{-5}$,  
$\mu_d(z)$ is of order $10^{-5}$ in the redshift range $1<z<100$. 
This property is independent of the unscreened and 
screened cosmological backgrounds, so the LLR constraint 
gives a tight restriction on the deviation of $d_{\rm GW}(z)$ 
from $d_{L}(z)$ in nonminimally coupled theories.

In this paper we did not study the evolution of scalar cosmological 
perturbations relevant to the observations of large-scale structures and 
weak lensing, but it is straightforward to do so by using the linear 
perturbation equations of motion derived in Refs.~\cite{Horn2,KT18,DKT}. 
In the unscreened cosmological background the upper limit of 
$Q$ is tightly constrained from the  LLR experiment, 
so the effective gravitational 
couplings felt by matter and light are close to $G_{\rm N}$ \cite{KT18}. 
In the screened background not only $\Omega_K$ but also 
$\Omega_{G_3}$ is generally much smaller than 1 at low redshifts, so it is expected that the gravitational 
interaction is not substantially modified from that in GR.
At the background level, the dark energy equations of state in the 
unscreened and screened cases exhibit some difference especially 
in the late cosmological epoch.
It will be of interest to place further constraints 
on the allowed parameter space of our theory 
by exploiting the observational 
data of cosmic expansion and growth histories.

\section*{Acknowledgments}
The author thanks Michele Maggiore for useful discussions.
The author is supported by the Grant-in-Aid for Scientific Research 
Fund of the JSPS No.~16K05359 and MEXT KAKENHI Grant-in-Aid 
for Scientific Research 
on Innovative Areas ``Cosmic Acceleration'' (No.\,15H05890). 


\end{document}